\documentclass[]{jfm}

\usepackage{graphicx}
\usepackage{amsmath}
\usepackage{natbib}
\usepackage{color}
\usepackage{array}
\usepackage{multirow}
\usepackage{xcolor}

\ifCUPmtlplainloaded \else
  \checkfont{eurm10}
  \iffontfound
    \IfFileExists{upmath.sty}
      {\typeout{^^JFound AMS Euler Roman fonts on the system,
                   using the 'upmath' package.^^J}%
       \usepackage{upmath}}
      {\typeout{^^JFound AMS Euler Roman fonts on the system, but you
                   dont seem to have the}%
       \typeout{'upmath' package installed. JFM.cls can take advantage
                 of these fonts,^^Jif you use 'upmath' package.^^J}%
      }
  \else
  \fi
\fi

\ifCUPmtlplainloaded \else
  \checkfont{msam10}
  \iffontfound
    \IfFileExists{amssymb.sty}
      {\typeout{^^JFound AMS Symbol fonts on the system, using the
                'amssymb' package.^^J}%
       \usepackage{amssymb}%
       \let\le=\leqslant  
         
      }{}
  \fi
\fi

\ifCUPmtlplainloaded \else
  \IfFileExists{amsbsy.sty}
    {\typeout{^^JFound the 'amsbsy' package on the system, using it.^^J}%
     \usepackage{amsbsy}}
    {}
\fi

\newsavebox{\astrutbox}
\sbox{\astrutbox}{\rule[-5pt]{0pt}{20pt}}

\newcommand{\sgn}{sgn}

\graphicspath{
{F:/Ph. D.,/}
}
\title[Spontaneous superharmonic internal wave generation by modal interactions]{Spontaneous superharmonic internal wave generation by modal interactions in uniform and nonuniform stratifications}

\author[Varma et. al.]
{
D\ls H\ls E\ls E\ls R\ls A\ls J\ns V\ls A\ls R\ls M\ls A\ls $^1$, V\ls A\ls M\ls S\ls I\ns K.\ns C\ls H\ls A\ls L\ls A\ls M\ls A\ls L\ls L\ls A\ls $^2$
\and \ns M\ls A\ls N\ls I\ls K\ls A\ls N\ls D\ls A\ls N\ns M\ls A\ls T\ls H\ls U\ls R\ls $^1$}

\affiliation{$^1$Department of Aerospace Engineering, Indian Institute of Technology Madras,
Chennai - 600036, India.\\
$^2$Department of Applied Mechanics, Indian Institute of Technology Delhi,
New Delhi - 110016, India.}

\begin{document}

\maketitle

\begin{abstract}
Internal tides and near-inertial internal waves in the ocean are well-recognized to play an important role in the global energy budget. Triadic resonance is one mechanism via which these internal waves dissipate their energy, often at locations away from their generation sites. In this paper, we perform a combined theoretical and numerical study of triadic resonance in internal wave modes in a finite-depth ocean with background rotation and an arbitrary stratification profile. The spatial evolution of the modal amplitudes within a resonant triad are first derived based on the requirement that the nonlinear solution at leading order cannot diverge. The amplitude evolution equations are then numerically solved for two different cases: (i) modes 1 and 2 at a specific frequency ($\omega_0$) in triadic resonance with the mode-1 superharmonic wave (frequency 2$\omega_0$) in a uniform stratification, and (ii) a self-interacting mode-3 at a specific frequency $\omega_0$ in triadic resonance with the mode-2 at frequency $2\omega_0$ in a nonuniform stratification representative of the ocean. Quantitative estimates of energy transfer rates within the resonant triad show that superharmonic wave generation resulting from modal interactions should be an important consideration alongside other triadic resonances like parametric subharmonic instability (PSI). Remarkably, in case (ii), the amplitude evolution equations suggest that any initial energy in mode-3 at frequency $\omega_0$ would get permanently transferred to mode-2 at frequency 2$\omega_0$. Direct numerical simulations are then performed to show the spontaneous generation of superharmonic internal waves resulting from modal interactions in the aforementioned two cases, and quantitatively validate the initial spatial evolution of the wave field predicted by the amplitude evolution equations. Furthermore, direct numerical simulations at off-resonant frequencies are used to identify the range of primary wave frequencies (around the resonant frequency) over which spontaneous superharmonic wave generation occurs. We conclude by giving estimates of the relative importance of superharmonic wave generation in the ocean, and provide motivation for future studies.
\end{abstract}

\section{Introduction}

\begin{figure}
\begin{center}
\includegraphics [width=\textwidth]{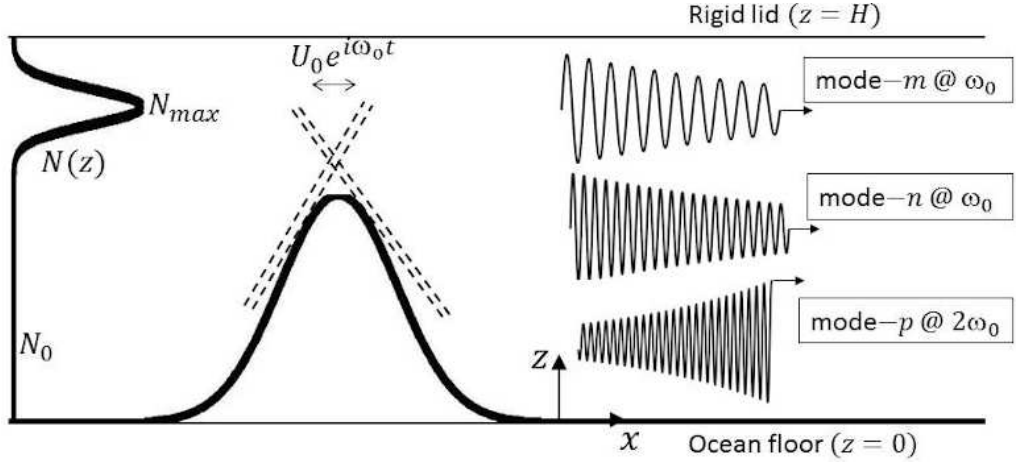}
\vspace{0.1cm}
\caption{Schematic showing the generation of internal wave beams by barotropic forcing (at frequency $\omega_0$) acting on a model ocean floor topography. The nonlinear spatial evolution of the internal wave beams, which are represented as a superposition of linear internal wave modes away from the topography, is the subject of the current study. The generation, and subsequent growth, of a superharmonic $2\omega_0$ internal wave mode-p as a result of its resonant interaction with modes m \& n at frequency $\omega_0$ are indicated (not drawn to scale) in the region to the right of the topography. A nonuniform stratification $N(z)$ with a pycnocline, representative of the ocean, is shown on the left. The ratio between the maximum stratification $N_{max}$ and the deep ocean value $N_0$ is unity for a uniformly stratified ocean.}
\label{cartoon1}
\end{center}
\end{figure}

Mechanisms that drive internal wave dissipation in the ocean are of paramount importance in understanding the global energy budget, the state of the ocean \& its spatio-temporal variability, and in improving the parametrization of internal wave driven mixing in general circulation models \citep{garrett_munk, MunkWunsch98, garrett2003, alford2016}. It is now well-established that a significant amount of energy is put into internal tides \citep{egbertray} and near-inertial motions \citep{alford2003}, which often propagate quite large distances both horizontally and vertically \citep{alford03, rainville2006, alford2016}. While internal tides are generated by barotropic forcing over ocean floor topography \citep{GarrettKunze07}, near-inertial internal waves are generated by winds acting on the ocean surface \citep{pollard1970, alford2016}. In this paper, we focus on one mechanism by which energy could be transferred from these internal tides or near-inertial internal waves into higher frequencies and other spatial scales, and potentially lead towards internal wave dissipation. Specifically, we study the generation of superharmonic internal waves resulting from a resonant interaction between internal wave modes.

Internal tide generation by barotropic forcing over continental shelves or deep ocean topography often occur in the form of internal wave beams \citep{GarrettKunze07}, i.e. a distribution of the overall energy into a series of vertical modes at the tidal frequency in the finite depth ocean. As shown in the schematic in figure \ref{cartoon1}, one of the goals of the current study is to highlight the importance of superharmonic internal wave generation due to resonant or near-resonant interaction between these modes at the tidal frequency. In a similar vein, the near-inertial internal wave field around their generation region can be represented as a summation over vertical modes at near-inertial frequencies \citep{gill, van2011, alford2016}. 

Triadic resonance \citep{leblond} is considered to be an important mechanism by which internal wave energy at a given frequency gets transferred to smaller spatial scales and other frequencies. For plane internal waves in a uniform stratification, the classical conditions of $\omega_1\pm\omega_2\pm\omega_3 = 0$ and ${\bf k_1} + {\bf k_2} + {\bf k_3} = 0$ are necessary to be satisfied by the wave components in a resonant triad. Here, $\omega_i(>0)$  and ${\bf k_i}$ are the frequency and wave vector of the $i^{th}$ wave in the resonant triad. Previous theoretical studies \citep{klostermeyer1991, sonmor1997} have even argued that resonant triad interactions are the basic mechanism behind every instability in a plane internal wave. A more recent theoretical study has also shown that several instabilities in a plane internal wave are related to triadic resonances at various orders \citep{ghaemsaidi2019}. An often-studied manifestation of these resonant triads is in the form of a primary wave forced at frequency $\omega_0$ generating two daughter waves at subharmonic frequencies ($\omega_1,\omega_2<\omega_0$) - we refer to such resonant triads as subharmonic resonant triads in the rest of this paper. In the limit of $\omega_1\approx\omega_2\approx\omega_0/2$, the generation of the subharmonic daughter waves is referred to as parametric subharmonic instability (PSI) \citep{staquet}. 

Numerical simulations \citep{Koudella_Staquet} and laboratory experiments \citep{Bourget_etal} have revealed the occurrence of subharmonic triadic resonance in plane internal waves, with both studies reporting comparisons with theoretical growth rates. The theoretical growth rates for plane internal waves in a uniform stratification are based on amplitude evolution equations derived using the method of multiple scales \citep{Bourget_etal,richet2018}. Subharmonic resonant triads, and thier associated amplitude evolution have also been studied in internal wave beams in an unbounded uniform stratification \citep{karimi_akylas, bourget2014, dauxois}. Furthermore, \cite{karimi2017} have investigated subharmonic triadic resonance in internal wave beams at twice the inertial frequency associated with the background rotation, again in an unbounded uniform stratification. In ocean-like nonuniform stratifications, the interaction between an internal wave beam and the pycnocline has also been reported to result in subharmonic triadic resonance \citep{gayen_sarkar}. Recently, \cite{saranraj}  used the method of multiple scales to derive the amplitude evolution equations for a resonant triad comprising internal wave packets in an unbounded weakly nonuniform stratifications with no background rotation.

With respect to internal wave modes, \cite{JFM2017} have theoretically shown the existence of resonant triads comprising two modes at a given frequency $\omega_0$ and a third mode at the superharmonic frequency $2\omega_0$ in uniform and nonuniform stratifications with background rotation. They further showed that ocean-like nonuniform stratifications containing a pycnocline support several more such resonant triads than a uniform stratification. \cite{JFM2017}, however, did not attempt to calculate the intra energy transfer rates within these resonant triads, which could be obtained by deriving the amplitude evolution equations. Previous studies have considered the temporal evolution of the wave amplitudes in resonant triads comprising internal wave modes in a uniform stratification with no background rotation, and have shown in laboratory experiments that complex wave fields containing different resonant triads emerge when individual standing modes of sufficiently strong amplitude are forced \citep{thorpe, mcewan, martin, mcewan1972, benielli1998}.

 \cite{liang2017} used the amplitude evolution equations for resonant triads comprising internal wave modes in a uniform stratification with no background rotation to show that PSI (daughter waves at half the primary wave frequency) does not necessarily represent the most unstable resonant triad that comprises a primary wave at frequency $\omega_0$ and two daughter waves at frequencies smaller than $\omega_0$. The subharmonic waves generated in an experimental study of a mode-1 internal wave in a uniform stratification also indicate that PSI with daughter waves at $\omega_0/2$ is not necessarily the most unstable resonant triad \citep{Joubaud_etal}. In the presence of background rotation, the numerical study of \cite{mackinnon_winters}  has shown the occurrence of PSI in mode-1 at twice the inertial frequency in a uniform stratification. The amplitude evolution for PSI in mode-1 at twice the inertial frequency in an ocean-like nonuniform stratification has been investigated by \cite{young_etal}. A recent numerical study by \cite{richet2018} has studied a similar problem in more detail, focusing on both the regimes of proapagation and evanescence for the daughter waves at the inertial frequency.     

\cite{sutherland} performed numerical simulations to show that individual modes in a nonuniform stratification can generate superharmonic disturbances. These superharmonic disturbances were not necessarily internal waves, and the focus of their study was not triadic resonance; rather, they provide a superharmonic generation mechanism based on weakly nonlinear interactions. Furthermore, the steady-state weakly nonlinear solution at the superharmonic frequency resulting from the non-resonant self-interaction of an isolated mode in ocean-like model nonuniform stratifications \citep{JFM2017, wunsch2017} suggest that the generation of superharmonic disturbances could be an important factor in the ocean. In this paper, we study the generation of superharmonic internal waves by resonant interaction between primary internal wave mode(s) at frequency $\omega_0$ in uniform and nonuniform stratifications with background rotation. Typically, in subharmonic triadic resonance, a finite amount of energy is put into a single primary wave, while the two subharmonic daughter waves have negligible energy to begin with. In contrast, our study concerns a superharmonic resonant triad, with the two primary waves at frequency $\omega_0$ containing a finite amount of initial energy, and hence potentially resulting in rapid growth of the single superharmonic daughter wave. The schematic in figure \ref{cartoon1}  depicts energy injection into two primary waves at tidal frequency by barotropic forcing on bottom topography, and the resulting spatial growth of a superharmonic internal wave.  

The main goals of this study are: (i) derivation of amplitude evolution equations for any resonant triad comprising internal wave modes in an arbitrary stratification profile with background rotation, (ii) highlight the significance of superharmonic wave generation as a result of resonant interaction between modes at a specific frequency, and (iii) perform direct numerical simulations (DNS) to validate the amplitude evolution equations, and investigate the superharmonic wave generation at off-resonant frequencies The rest of the paper is organized as follows. Section \ref{sec:theory} presents the derivation of the amplitude evolution equations. In section \ref{sec:numerics}, we present the details of DNS, and also motivate the specific cases we study numerically. Section \ref{sec:numerics} also includes a description of the post-processing of the flow fields obtained from DNS. Section \ref{sec:results} presents our theoretical and DNS results, which are then summarized in the beginning of section \ref{sec:disc}. We conclude by highlighting the relative importance of our results, and also discuss the scope for future studies.      

\section{Theory}
\label{sec:theory}

\subsection{Amplitude Evolution Equations}
\label{sec:AEE}
The fully nonlinear governing equations for two-dimensional (2D), incompressible, inviscid flow on the $f-$plane under the Boussinesq approximation are~\citep{leblond}
\begin{gather}
\frac{\partial^2}{\partial t^2}(\nabla^{2}\psi) + f^{2}\frac{\partial^{2}\psi}{\partial z^2} = \frac{g}{\bar{\rho}}\frac{\partial}{\partial x}[J(\psi,\rho)] - \frac{\partial}{\partial t}[J(\psi,\nabla^{2}\psi)] + f \frac{\partial}{\partial z}[J(\psi,v)], \label{psi_eqn} \\
\frac{\partial \rho}{\partial t} = -J(\psi,\rho), \label{rho_eqn}\\
\frac{\partial v}{\partial t} - f \frac{\partial \psi}{\partial z} = -J(\psi,v) , \label{v_eqn}
\end{gather}
where $x$, $z$ are the horizontal and vertical spatial coordinates, respectively, and $t$ is time. $\psi(x,z,t)$, $v(x,z,t)$ and $\rho(x,z,t)$ are the stream function, $y-$component of velocity and density, respectively, with the velocity components in the $xz$-plane given by $(u,w) = (-\partial\psi/\partial z,\partial\psi/\partial x)$. $\bar{\rho}$ is a reference constant density, $f$ the Coriolis frequency and ${\bf g} = -g{\bf \hat{e}_z}$ ($g>0$) the acceleration due to gravity ($\bf \hat{e}_z$ is the unit vector along the positive $z-$axis). The Jacobian operator $J$ is defined as $J(A,B) = (\partial A/\partial x)(\partial B/\partial z) - (\partial B/\partial x)(\partial A/\partial z)$, and $\nabla^2 = \partial^2/\partial x^2 + \partial^2/\partial z^2$. The no-normal flow boundary condition at the horizontal boundaries of the fluid of depth $H$ is specified as $\psi(x,z=0,t) = \psi(x,z=H,t) = 0$, where the ocean floor at $z=0$ is assumed flat and the free surface at $z=H$ is modelled as a rigid lid.

In the framework of regular perturbation expansion, we seek solutions to equations (\ref{psi_eqn})-(\ref{v_eqn}) of the form
\begin{gather}
(\psi,v,\rho) = (\psi_{0},v_{0},\rho_{0}) + \epsilon(\psi_{1},v_{1},\rho_{1}) + \epsilon^2(\psi_{2},v_{2},\rho_{2}) + \cdot\cdot\cdot \;,
\label{perturbation_expansion}
\end{gather}
where the small parameter $\epsilon$ quantifies the relative magnitude of the nonlinear terms in the governing equations. The quiescent stably stratified background state is described by the solution at $\mathcal{O}(\epsilon^0)$: $\psi_0 = 0$, $v_0 = 0$, $\rho_0(z) = \bar{\rho} - (\bar{\rho}/g)\int_0^zN^2(z)dz$, where $N(z)$ is the vertical profile of the buoyancy frequency.

We assume the existence of a slow horizontal coordinate $X = \epsilon x$ over which internal wave amplitudes at $\mathcal{O}(\epsilon)$ vary. Following the classical method of multiple scales \citep{nayfeh_book}, the horizontal spatial derivatives in the governing equations (\ref{psi_eqn})-(\ref{v_eqn}) are written as
\begin{gather}
\frac{\partial}{\partial x} = \frac{\partial}{\partial x} + \epsilon\frac{\partial}{\partial X}, \label{tau1}\\
\frac{\partial^2}{\partial x^2} = \frac{\partial^2}{\partial x^2} + 2\epsilon\frac{\partial^2}{\partial x \partial X} + \epsilon^2\frac{\partial^2}{\partial X^2}.\label{tau2}
\end{gather}
While it is common in the literature \citep{Bourget_etal,richet2018} to consider slow temporal evolution of the wave amplitudes in a resonant triad, our assumption of a wave field with slow spatial evolution for the amplitudes is better suited to describe scenarios like in figure \ref{cartoon1}, where constant-amplitude primary waves at a fixed frequency are continuously forced at the fixed location of the generation site. 

Substituting the solution forms in equation (\ref{perturbation_expansion}) into equations (\ref{psi_eqn})-(\ref{v_eqn}), and using equations (\ref{tau1})-(\ref{tau2}), the governing equations at $\mathcal{O} (\epsilon^0)$, $\mathcal{O} (\epsilon^1)$ and $\mathcal{O} (\epsilon^2)$ are obtained as shown in equations (\ref{eps0})-(\ref{eps2}) of appendix \ref{eps_eqn}. Substituting $\psi_0 = 0$, $v_0 = 0$ and $\rho_0(z) = \bar{\rho} - (\bar{\rho}/g)\int_0^zN^2(z)dz$, the governing equations at $\mathcal{O} (\epsilon^1)$ reduce to
\begin{gather}
\frac{\partial^2}{\partial t^2}(\nabla^{2}\psi_{1}) + f^{2}\frac{\partial^{2}\psi_{1}}{\partial z^{2}} +  N^{2}\frac{\partial^{2}\psi_{1}}{\partial x^{2}} = 0,  \label{psi1_eqn}\\
\frac{\partial \rho_{1}}{\partial t} = \frac{\bar{\rho}}{g}N^2\frac{\partial \psi_1}{\partial x}, \label{rho1_eqn} \\
\frac{\partial v_{1}}{\partial t} = f\frac{\partial \psi_1}{\partial z}, \label{v1_eqn}
\end{gather}
with $\psi_1$ satisfying the boundary conditions $\psi_1(x,z=0,t) = \psi_1(x,z=H,t) = 0$. Equations (\ref{psi1_eqn})-(\ref{v1_eqn}) coincide with the well-known linear internal wave equations.

We assume $(\psi_1,\rho_1,v_1)$ to be described by a superposition of three internal wave modes that form a resonant triad. Specifically, they take the following form
\begin{gather}
\psi_1 =  \sum_{j=1}^{3}\frac{1}{2}\Psi_j(X)\phi_j(z)e^{i(k_j x-\omega_j t)} + \frac{1}{2}\Psi^{*}_j(X)\phi_j(z)e^{-i(k_j x-\omega_j t)},\label{psi1_sol}\\
\rho_1 =  \sum_{j=1}^{3}\frac{1}{2}R_j(X)\rho_j(z)e^{i(k_j x-\omega_j t)} + \frac{1}{2}R^{*}_j(X)\rho_j(z)e^{-i(k_j x-\omega_j t)},\label{rho1_sol}\\
v_1 =  \sum_{j=1}^{3}\frac{1}{2}C_j(X)v_j(z)e^{i(k_j x-\omega_j t)} + \frac{1}{2}C^{*}_j(X)v_j(z)e^{-i(k_j x-\omega_j t)},\label{v1_sol}
\end{gather}
where the superscript $*$ denotes complex conjugate. The frequencies (assumed real \& positive) and horizontal wavenumbers (assumed real) satisfy the triadic resonance conditions: $\omega_1+\omega_2 = \omega_3$ and $k_1+k_2 = k_3$ \citep{leblond}.
The real vertical mode shapes $\phi_j(z)$ satisfy
\begin{gather}
\frac{d^{2}\phi_{j}}{dz^{2}} + \frac{k_{j}^{2}(N^2 - \omega^{2}_{j})}{\omega^{2}_{j} - f^2}\phi_j = 0, \label{phi_equation}
\end{gather}
along with the boundary conditions $\phi_j(z=0) = \phi_j(z=H) = 0$. Additionally, for triadic resonance, the mode shapes are required to satisfy a non-orthogonality condition similar to what is derived in \cite{JFM2017} for the special case of $\omega_1 = \omega_2$. This additional condition for the vertical mode shapes causes the non-existence of a steady-state (with non-zero amplitudes for all the three waves in the resonant triad) for the amplitude evolution equations we derive in this subsection. For a uniform stratification, it is equivalent to the requirement that the mode numbers $m_j$ associated with $\phi_{j}(z)$ satisfy $m_3 = \vert m_2 - m_1 \vert$. In contrast, in a non-uniform stratification, the conditions $ k_1 + k_2 = k_3$ and $\omega_1 + \omega_2 = \omega_3$ are likely sufficient to ensure that the additional condition on the vertical mode shapes required for triadic resonance is satisfied \citep{JFM2017}. Substituting equations (\ref{psi1_sol})-(\ref{v1_sol}) in equations (\ref{rho1_eqn})-(\ref{v1_eqn}) gives the following relations:
\begin{gather}
R_{j}(X)\rho_{j}(z) = (\frac{N^2\bar{\rho}}{g})(\frac{-k_j}{\omega_j})\phi_{j}(z)\Psi_{j}(X),\label{Rj}\\
C_j(X)v_{j}(z) = i\frac{f}{\omega_j}\frac{d\phi_{j}}{dz}\Psi_j(X),\label{Cj}
\end{gather}
where $\Psi_j$, $R_j$ and $C_j$ are the complex amplitudes of the $j^{th}$ wave for $\psi_1$, $\rho_1$ and $v_1$, respectively, and the corresponding real vertical mode shapes are $\phi_j(z)$, $\rho_j(z)$ and $v_j(z)$. For convenience, the vertical mode shapes are defined such that
\begin{gather}
\frac{\bar{\rho}(\omega_{j}^{2} -f^2)}{2\omega_{j}k_j}\int_0^H \big(\frac{d\phi_j}{dz}\big)^2 dz = 1, \label{norm}
\end{gather} 
corresponding to unit energy flux in the $j^{th}$ internal wave if $|\Psi_j| = 1$. In other words, $|\Psi_j|^2$ is the energy flux in the $j^{th}$ wave of the resonant triad. 

In the ocean, internal tide generation and/or scattering at the bottom topography excite at least a few internal wave modes at the tidal frequency, with a significant fraction of the total energy being in the low modes. In the current study, this low-mode internal wave field is assumed to contain one or two of the three internal waves in the resonant triad we consider in equations (\ref{psi1_sol}) - (\ref{v1_sol}). For example, two modes at the semidiurnal frequency could represent the primary waves, and they could be in triadic resonance with a superharmonic secondary wave \citep{JFM2017}. Alternately, one mode at the semidiurnal frequency could be the primary wave, which is in triadic resonance with two secondary waves  at subharmonic frequencies. The relative importance of these various resonant triads which comprise one or two of the waves in the existing linear internal wave field is the focus of our current study.

Equation (\ref{psi_eqn}), upon substituting $\psi_0 = 0$, $v_0 = 0$ and $\rho_0(z) = \bar{\rho} - (\bar{\rho}/g)\int_0^zN^2(z)dz$, is rewritten at
$\mathcal{O}(\epsilon^2)$ as
\begin{align}
\frac{\partial^2}{\partial t^2}(\nabla^{2}\psi_{2}) + f^{2} \frac{\partial^{2}\psi_{2}}{\partial z^{2}} + N^{2} \frac{\partial^{2}\psi_{2}}{\partial x^{2}} = &\frac{g}{\bar{\rho}}\frac{\partial}{\partial x}[J(\psi_{1},\rho_{1})] - \frac{\partial}{\partial t}[J(\psi_{1},\nabla^{2}\psi_{1})]\nonumber \\
&+ f \frac{\partial}{\partial z}[J(\psi_1,v_1)] - 2(\frac{\partial^{2}}{\partial t^{2}} + N^2)\frac{\partial^{2}\psi_1}{\partial x \partial X},
\label{psi2_eqn}
\end{align}
where the terms involving the $\mathcal{O}(\epsilon^1)$ wave field $(\psi_1,\rho_1,v_1)$ serve as forcing terms for the $\mathcal{O}(\epsilon^2)$ wave field $\psi_2$. Using the expressions (\ref{psi1_sol}) - (\ref{v1_sol}) and (\ref{Rj}) - (\ref{Cj}), the right hand side ($R$) of equation (\ref{psi2_eqn}) is
\begin{gather}
R = \sum_{p=1}^{3}\sum_{q=1}^{3}\frac{1}{4}\Bigg[A_{pq}(z)\Psi_p\Psi_q e^{i[(k_p+k_q)x-(\omega_p + \omega_q)t]} + B_{pq}(z) \Psi_p \Psi^{*}_q e^{i[(k_p-k_q)x-(\omega_p-\omega_q)t]}\Bigg]\nonumber\\
- i\sum_{j=1}^{3}k_{j}\frac{d\Psi_j}{dX}\phi_j(z)[N^2 - \omega
^{2}_{j}] e^{i(k_j x-\omega_j t)} + c.c.,
\label{RHS_simplified}
\end{gather}
where $A_{pq}(z)$ and $B_{pq}(z)$ are given by
\begin{align}
A_{pq}(z) = &\frac{k_q}{\omega_q}(k_p+k_q)\Bigg[k_p N^{2}\phi_{p}\frac{d\phi_q}{dz} + k_p\frac{d(N^2)}{dz}\phi_{p}\phi_{q} - k_qN^{2}\frac{d\phi_p}{dz}\phi_{q}\Bigg]\nonumber\\
&-(\omega_{p} + \omega_{q})\Bigg[k_{p}\phi_{p}(\frac{d^{3}\phi_q}{dz^3}-k^{2}_{q}\frac{d\phi_q}{dz}) - k_{q}\frac{d\phi_p}{dz}(\frac{d^{2}\phi_q}{dz^2}-k^{2}_{q}\phi_q)\Bigg]\nonumber\\
&-\frac{f^{2}}{\omega_{q}}\Bigg[k_{p}\frac{d\phi_p}{dz}\frac{d^{2}\phi_q}{dz^2} + k_p\phi_{p}\frac{d^{3}\phi_q}{dz^3} - k_{q}\frac{d^{2}\phi_p}{dz^2}\frac{d\phi_q}{dz} -k_{q}\frac{d\phi_p}{dz}\frac{d^{2}\phi_q}{dz^{2}}\Bigg], \label{Apq}\\
B_{pq}(z) = &\frac{k_q}{\omega_q}(k_p-k_q)\Bigg[k_pN^{2}\phi_{p}\frac{d\phi_q}{dz} + k_p\frac{d(N^2)}{dz}\phi_{p}\phi_{q} + k_qN^{2}\frac{d\phi_p}{dz}\phi_{q}\Bigg]\nonumber\\
&-(\omega_{p} - \omega_{q})\Bigg[k_{p}\phi_{p}(\frac{d^{3}\phi_q}{dz^3}-k^{2}_{q}\frac{d\phi_q}{dz}) + k_{q}\frac{d\phi_p}{dz}(\frac{d^{2}\phi_q}{dz^2}-k^{2}_{q}\phi_q)\Bigg]\nonumber\\
&+\frac{f^{2}}{\omega_{q}}\Bigg[k_{p}\frac{d\phi_p}{dz}\frac{d^{2}\phi_q}{dz^2} + k_p\phi_{p}\frac{d^{3}\phi_q}{dz^3} + k_{q}\frac{d^{2}\phi_p}{dz^2}\frac{d\phi_q}{dz} + k_{q}\frac{d\phi_p}{dz}\frac{d^{2}\phi_q}{dz^2}\Bigg].\label{Bpq}
\end{align}

The expression for $R$ in equation (\ref{RHS_simplified}), upon incorporating the triadic resonance conditions of $\omega_1 + \omega_2 = \omega_3$ and $k_1 + k_2 = k_3$, is re-written as
\begin{equation}
R  = \sum_{j=1}^{3}Q_{j}(z)e^{i(k_j x-\omega_j t)} +  NRT,\label{resonant_terms}
\end{equation}
where NRT refers to the non-resonant terms that are not at any of the frequencies ($\omega_1,\omega_2$ or $\omega_3$) of the resonant triad making up the $\mathcal{O}(\epsilon)$ wave field. In equation (\ref{resonant_terms}), $Q_j(z)$ for $j = 1,2,3$ are
\begin{gather}
Q_1(z) = -i[N^2 - \omega
^{2}_{1}]k_1\phi_1(z)\frac{d\Psi_1}{dX} + \frac{1}{4}[B_{23}(z)+B_{32}(z)]\Psi^{*}_2\Psi_3,\label{Q1}\\
Q_2(z) = -i[N^2 - \omega
^{2}_{2}]k_2\phi_2(z)\frac{d\Psi_2}{dX} + \frac{1}{4}[B_{13}(z)+B_{31}(z)]\Psi^{*}_1\Psi_3,\label{Q2}\\
Q_3(z) = -i[N^2 - \omega
^{2}_{3}]k_3\phi_3(z)\frac{d\Psi_3}{dX} + \frac{1}{4}[A_{12}(z)+A_{21}(z)]\Psi_1\Psi_2.\label{Q3}
\end{gather}

In the event of a triadic resonance resulting from self-interaction of a mode in a nonuniform stratification \citep{JFM2017}, i.e. $\omega_1 = \omega_2 = \omega_3/2$ and $k_1 = k_2 = k_3/2$, two additional terms corresponding to $p=q=1$ and $p=q=2$ in equation (\ref{RHS_simplified}) will also contribute to equation (\ref{Q3}). Thus, equation (\ref{Q3}) gets modified to
\begin{gather}
Q_3(z) = -i[N^2 - \omega^{2}_{3}]k_3\phi_3(z)\frac{d\Psi_3}{dX} + \frac{1}{4}[A_{11}(z)+A_{12}(z)+A_{21}(z)+A_{22}(z)]\Psi_1\Psi_2.\label{Q3_mod}
\end{gather}
Furthermore, with $\Psi_1 = \Psi_2$, $2\Psi_1$ is the amplitude of the self-interacting mode, and the corresponding energy flux is 4$|\Psi_1|^2$.

The solution to equation (\ref{psi2_eqn}), corresponding to the resonant terms in $R$ shown in equation (\ref{RHS_simplified}), is sought as
\begin{gather}
\psi_2  = \sum_{j=1}^{3}h_{j}(z)e^{i(k_j x-\omega_j t)} + c.c.,\label{psi2_sol}
\end{gather}
where $h_j(z)$ is complex in general. Substituting equation (\ref{psi2_sol}) in equation (\ref{psi2_eqn}), and equating the coefficients of $e^{i(k_j x-\omega_j t)}$ gives the governing equation for $h_j(z)$ as
\begin{gather}
\frac{d^{2} h_{j}}{dz^{2}} + \frac{k_{j}^{2}(N^2 - \omega^{2}_{j})}{\omega^{2}_{j} - f^2} h_j = C_j,\label{hj_equation}
\end{gather}
where $C_{j}(z) = -Q_{j}(z)/(\omega^{2}_{j} - f^2)$. No-normal flow boundary conditions at $z=0$ and $z=H$ require $h_j(z=0) = h_j(z=H) = 0$. Equation (\ref{hj_equation}) is essentially the same as the governing equation (\ref{phi_equation}) for the mode shapes $\phi_{j}(z)$, but now with a non-zero forcing term $C_{j}(z)$ on the right hand side. In other words, $\phi_{j}(z)$ represents the homogeneous solution to equation (\ref{hj_equation}).

The particular solution to equation (\ref{hj_equation}) would diverge unless the orthogonality condition
\begin{gather}
\int_{0}^{H}\phi_{j}(z)C_{j}(z)dz=0 \label{h_ortho}
\end{gather}
is satisfied. In other words, divergence of $h_j(z)$ due to triadic resonance \citep{JFM2017} would be avoided if the condition (\ref{h_ortho}) is satisfied. The amplitude evolution equations are hence derived by enforcing equation (\ref{h_ortho}) for each of $j = 1,2,3$, to give
\begin{gather}
\frac{d\Psi_1}{dX} = i\alpha_1\Psi^{*}_2\Psi_3,\label{amp1_eqn}\\
\frac{d\Psi_2}{dX} = i\alpha_2\Psi^{*}_1\Psi_3,\label{amp2_eqn}\\
\frac{d\Psi_3}{dX} = i\alpha_3\Psi_1\Psi_2,\label{amp3_eqn}
\end{gather}
where the constants $\alpha_1$, $\alpha_2$ and $\alpha_3$ are given by
\begin{gather}
\alpha_1 = \frac{-1}{4k_1}\frac{\int_{0}^{H}\phi_{1}(z)[B_{23}(z)+B_{32}(z)]dz}{\int_{0}^{H}\phi^{2}_{1}(z)[N^2 - \omega^{2}_{1}]dz},\label{alpha1}\\
\alpha_2 = \frac{-1}{4k_2}\frac{\int_{0}^{H}\phi_{2}(z)[B_{31}(z)+B_{13}(z)]dz}{\int_{0}^{H}\phi^{2}_{2}(z)[N^2 - \omega^{2}_{2}]dz},\label{alpha2}\\
\alpha_3 = \frac{-1}{4k_3}\frac{\int_{0}^{H}\phi_{3}(z)[A_{12}(z)+A_{21}(z)]dz}{\int_{0}^{H}\phi^{2}_{3}(z)[N^2 - \omega^{2}_{3}]dz}.
\label{alpha3}
\end{gather}
The amplitude evolution equations (\ref{amp1_eqn})-(\ref{amp3_eqn}) satisfy energy conservation in the form of
\begin{equation}
\frac{d}{dX}\bigg[\sgn(k_1)|\Psi_1|^2 + \sgn(k_2)|\Psi_2|^2 + \sgn(k_3)|\Psi_3|^2)\bigg] = 0, 
\end{equation} 
where $\sgn(k_i)$ represents the sign of $k_i$ and hence the direction of propagation of the $i^{th}$ wave.

It is noteworthy that $\alpha_1$, $\alpha_2$ and $\alpha_3$ are non-zero since the three waves in $\psi_1$ (equation \ref{psi1_sol}) form a resonant triad. Specifically, \cite{JFM2017} showed that three constant amplitude waves (i.e. no spatial dependence for $\Psi_1$, $\Psi_2$ and $\Psi_3$) with their frequencies and horizontal wavenumbers satisfying $\omega_1 + \omega_2 = \omega_3$ and $k_1 + k_2 = k_3$ represent a resonant triad only if $\alpha_1$, $\alpha_2$ and $\alpha_3$ are non-zero. In the current study, the weak spatial dependence (i.e variatian in $X$) in the amplitudes $\Psi_1$, $\Psi_2$ and $\Psi_3$ are such that diverging steady-state solutions due to triadic resonance \citep{JFM2017} is avoided. In the case of a self-interacting mode in triadic resonance with a superharmonic mode, $\alpha_3$ is twice the value given by the expression in equation (\ref{alpha3}), owing to the additional terms shown in equation (\ref{Q3_mod}). Additionally, for self-interaction, we prescribe $\Psi_1 = \Psi_2$, $\omega_1 = \omega_2$, $k_1 = k_2$ and $\phi_1 = \phi_2$, where $2\Psi_1$ represents the streamfunction amplitude of the self-interacting mode.    

\subsection{Off-resonance steady state solutions}
\label{offres}
Weakly nonlinear steady-state solutions resulting from an interaction between modes $m$ \& $n$ at frequency $\omega_0$ in the linear wave field were studied by Varma \& Mathur (2017). In this framework, the wave field at order $\epsilon^1$ is written as:
\begin{gather}
\psi_1 =  \Psi_1\phi_1(z)\cos(k_{1}x-\omega_0 t + \gamma_1) + \Psi_2\phi_2(z)\cos(k_{2}x-\omega_0 t + \gamma_2),\label{psi1_steady}
\end{gather}
where $\Psi_1$ and $\Psi_2$ are real constants with no dependence on space or time. The solution at order $(\epsilon^2)$ then turns out to be \citep{JFM2017} 
\begin{gather}
\psi_2 = \bar{h}_{12}(z)\cos[(k_1+k_2)x-2\omega_0 t + \gamma_1 +\gamma_2]
+ \bar{g}_{12}(z)\cos[(k_1-k_2)x + \gamma_1 -\gamma_2], \label{psi2_steady}
\end{gather}
comprising a superharmonic wave at frequency $2\omega_0$ (and horizontal wavenumber $k_1+k_2$) and a time-independent mean flow of horizontal wavenumber $k_1-k_2$ (with vertical structure $\bar{g}_{12}(z)$). The vertical structure, $\bar{h}_{12}(z)$ of the superharmonic wave is governed by
\begin{gather}
\frac{d^2\bar{h}_{12}}{dz^2} + (k_1+k_2)^2 \frac{N^2 - 4\omega^2}{4\omega^2-f^2}\bar{h}_{12} = \Psi_1\Psi_2\bar{C}_{12},
 \label{h_mn}
\end{gather}
with $\bar{C}_{12} = -(A_{12}+A_{21})/(4\omega^2-f^2)$, where $A_{12}$ is as defined in equation (\ref{Apq}). $\bar{h}_{12}(z)$  satisfies the boundary conditions $\bar{h}_{12}(z=0) = \bar{h}_{12}(z=H) = 0$. If modes $m$ \& $n$ at frequency $\omega_0$ are in triadic resonance with the superharmonic wave in equation (\ref{psi2_steady}), the resulting $\bar{h}_{12}$ would diverge \citep{JFM2017}. Away from the triadic resonance, equation (\ref{h_mn}) can be numerically solved to obtain a finite solution, with the magnitude of $\bar{h}_{12}$ diverging towards infinity as we approach triadic resonance. In the current study, a plot of the magnitude of $\bar{h}_{12}$ (defined as $\smash{h_{max} = \displaystyle\max}[\bar{h}_{12}(z)]$) in the neighbourhood of triadic resonance parameters provides a reference using which the  near-resonant behaviour observed in numerical simulations is interpreted. 

\section{Numerical simulations}
\label{sec:numerics}

Direct numerical simulations are performed to (i) demonstrate the spontaneous generation of the theoretically predicted superharmonic internal wave due to resonant interaction between internal wave modes, (ii) quantitatively validate the initial spatial growth of the superharmonic wave estimated from the amplitude evolution equations derived in section \ref{sec:AEE} , (iii) investigate off-resonance superharmonic generation, which is outside the scope of the theory presented in section \ref{sec:AEE}. The aforementioned goals are achieved for two representative scenarios: (a) modes 1 \& 2 interacting in a uniform stratification, and (b) self-interaction of mode 3 in a nonuniform stratification. In the rest of this section, we describe the numerical framework, list out all the numerical simulations performed, and present the methods used to analyze the output data from the numerical simulations.

\subsection{ The SOMAR framework}

Two dimensional numerical simulations are performed using SOMAR \citep{SOMAR,vamsi2017}, which is an adaptive mesh numerical framework.  SOMAR solves a set of conservation equations for momentum and buoyancy on a composite grid consisting of several levels, each at different resolution.  This increases efficiency and allows the resolution of a wide range of scales with a reduced computational cost when compared to a traditional single-level solver. In the current study, however we consider a single level grid at a uniform resolution in both horizontal and vertical directions. For details about the adaptive mesh capabilities of SOMAR, see \cite{SOMAR,vamsi2017}. SOMAR uses second-order, central finite differences for the spatial discretization of viscous and diffusion terms, computed semi-implicitly for time advancement.  Advection terms are computed using the finite volume Piecewise-Parabolic Method (PPM) of \cite{Colella1984}.

\subsection{Governing equations}
\noindent Two dimensional Navier-Stokes equations are solved on cartesian grids. In SOMAR, the buoyancy field is split into the background $\overline{b}(z)$ and perturbation $b^*(x,y,z,t)$ fields:
\begin{gather}\label{F01}
    b(x,y,z,t) = \overline{b}(z)+b^*(x,y,z,t).
\end{gather}
The background buoyancy is given by $\overline{b}(z) = -\int_0^z N^2 dz$, where $N(z)$ is the background buoyancy frequency profile. The governing equations are given by
\begin{gather}
{\mathbf {\nabla}}\cdot  {\mathbf u} = 0, \label{divu}\\
\frac{D  {\mathbf u}  }{D t} = - {\mathbf {\nabla}} p ^* - b^*  \mathbf{ \hat{e}_z} + \nu \nabla ^2  \mathbf u,\label{NS} \\
\frac{D b ^*}{D t} =  \kappa {\mathbf \nabla} ^2 b ^*  + w  N^2.\label{bp}
\end{gather}

\begin{figure}
\centering
\includegraphics[width=0.95\textwidth]{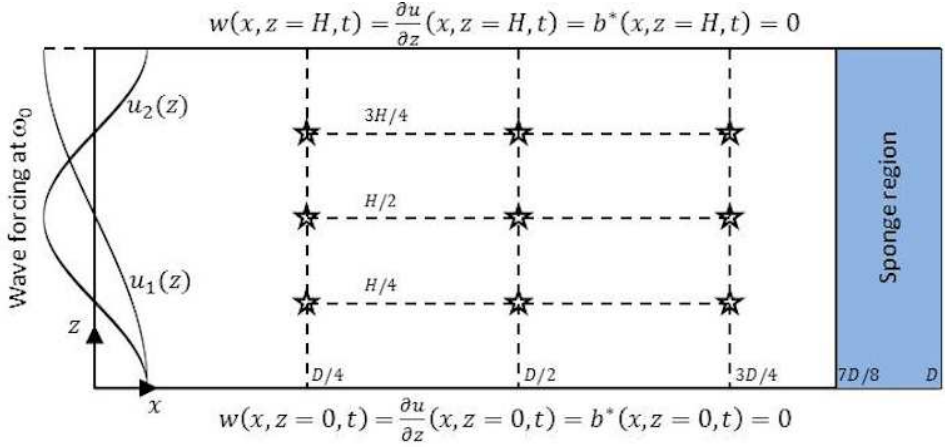}
\caption{Schematic of the computational set-up for the numerical simulations. Two specific internal wave modes $m$ \& $n$ at frequency $\omega_0$ are forced at $x=0$, and are allowed to evolve over the domain length $D$. A sponge layer of width $D/8$ at the right end of the domain ensures that no wave energy gets reflected back into the computational domain. The boundary conditions at $z = 0$ and $z=H$ are indicated next to the corresponding boundaries. The "star" symbols denote the spatial locations where the Fourier transform of the flow field are shown in figures 5, 9, 12 and 16.}
\label{cartoon2}
\end{figure}

Here, ${\mathbf u}=(u,w)$ represents the velocity vector with cartesian velocity components $u$ and $w$ in the $x$ (horizontal) and $z$ (vertical) directions, respectively. $\mathbf{\hat{e}_z}$ represents the unit vector in the vertical direction, $p ^*$ is the pressure deviation from the background hydrostatic pressure distribution, $\nu$ is the molecular viscosity, $\kappa$ is the molecular diffusivity. In all the numerical simulations considered in this study, the molecular viscosity and diffusivity values are taken to be $1 \times 10^{-6} m^2/s$. The governing equations in (\ref{divu})-(\ref{bp}) are numerically solved on a two-dimensional computational domain of length $D$ and height $H$. A schematic of the computational domain is shown in figure \ref{cartoon2}.

It is noteworthy that no explicit noise is added in our numerical simulations. In other words, the growth of any superharmonic waves is spontaneous and does not rely on any initial energy being put into them. The potentially important role of a prescribed background noise (of different amplitudes) would be investigated in a future study.  

\subsection{Boundary conditions}
At the left boundary ($x=0$), a combination of mode-m and mode-n waves at the forcing frequency $\omega_0$ are specified as a function of time. Specifically, the boundary conditions at $x=0$ are: \\
\begin{gather}
u(x=0,z,t) = (\frac{W_1}{k_1}\frac{d\phi_1}{dz} + \frac{W_2}{k_2}\frac{d\phi_2}{dz})\cos\omega_{0}t,\label{scalings1}\\
w(x=0,z,t) = (W_{1}\phi_1(z) + W_{2}\phi_2(z))\sin\omega_{0}t\label{scalings2}\\
b^*(x=0,z,t) = \frac{N^2}{\omega_0} (W_{1}\phi_1(z) + W_{2}\phi_2(z))\cos\omega_{0}t,\label{scalings3}
\end{gather}
where $W_1$ and $W_2$ are the vertical velocity modal amplitudes at the forcing (primary) frequency $\omega_0$. At the right boundary, a sponge forcing \citep{sponge_ref} is applied to absorb the internal wave energy. At the top ($z=H$) and bottom ($z=0$) boundaries, no-normal flow ($w=0$) and free-slip ($\partial u/\partial z = 0$) boundary conditions are enforced. The schematic in figure \ref{cartoon2} provides a summary of the boundary conditions. 

\subsection{Simulation cases: parameters, domain and grid resolution}
\label{methodology}
As mentioned in the beginning of this section, we perform numerical simulations for representative cases in uniform and nonuniform stratifications. 

\subsubsection{Uniform stratification}
\label{u_cases}
\begin{figure}
\centering
\includegraphics[width=1\textwidth]{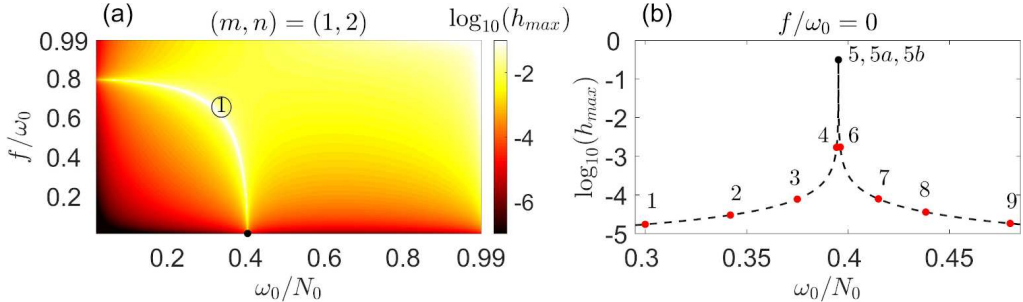}
\caption{(a) Steady-state amplitude of the superharmonic wave at $2\omega_0$ (resulting from interaction between mode-1 \& mode-2 at $\omega_0$, see equation  \ref{psi2_steady}) as a function of $\omega_0/N_0$ and $f/\omega_0$ in a uniform stratification (the encircled number denotes the mode number of the superharmonic wave on the triadic resonance curve), (b) Variation of $h_{max}$ (in log scale) with $\omega_0/N_0$ around the triadic resonance point (indicated by the solid black circle) for $f/\omega_0 = 0$. The red solid circles in (b) denote the off-resonance frequencies for which numerical simulations were performed, with the numbers next to them denoting the case numbers specified in table \ref{tab:Uniform stratification}. }
\label{steady_uniform}
\end{figure}

Here, we consider the interaction between modes 1 \& 2 in a uniform stratification $N_0$. For this case, \cite{JFM2017} showed the  existence of a unique curve on the $\omega_0/N_0-f/\omega_0$ plane on which triadic resonance occurs. A plot of the superharmonic wave amplitude based on the steady state theory (section \ref{offres}) reveals the curve along which modes 1 \& 2 at frequency $\omega_0$ resonantly interact to result in the generation of an infinitely large mode-1 internal wave at frequency 2$\omega_0$ (figure \ref{steady_uniform}a). Numerical simulations are performed at and around the triadic resonance curve in the limit of $f = 0$ (no background rotation), i.e. a close neighbourhood of the solid black circle in figure \ref{steady_uniform}(a). As shown in figure \ref{steady_uniform}(b), the steady state amplitude of mode-1 at frequency $2\omega_0$ (section \ref{offres}) diverges at $\omega_0/N_0 = 0.3953$ as a result of resonant interaction between modes 1 \& 2 at frequency $\omega_0$. The amplitude evolution equations derived in section \ref{sec:AEE} provide a non-diverging non-steady-state evolution of the resonant triad at these resonant frequencies. The numerical simulation in which primary wave modes 1 \& 2 at frequency $\omega_0/N_0 = 0.3953$ are forced at the left end of the computational domain is denoted as case 5, and is used to validate the amplitude evolution equations derived in section \ref{sec:AEE}. Cases 5a and 5b correspond to the same primary wave frequency as case 5, but with different domain length and primary wave forcing amplitudes, respectively. Cases 1-4 and 5-9 explore primary wave frequencies that are smaller and larger than the resonant frequency of $\omega_0/N_0 = 0.3953$, respectively. It is noteworthy that the amplitude evolution equations derived in section \ref{sec:AEE} are, strictly speaking, valid only at resonant frequencies. Numerical simulations at the off-resonant frequencies are therefore required to identify the range of frequencies over which triadic resonance-like behaviour occurs.  

\begin{table}
  \centering
  \renewcommand{\arraystretch}{1.25}
  \setlength{\tabcolsep}{4pt}
  \begin{tabular}{ c  c  c  c  c  c  c  c  c p{1cm} }
    Case & $\omega_0/N_0$  & $\lambda_1 = 2\pi/k_1(m)$  & $W_1(m/s)$  & $W_2(m/s)$ & $D/\lambda_{1}$  \\ \hline  
    1 & 0.3 & 24167 & -5.5154e-03 & -3.6769e-03 & 5 \\ 
    2 & 0.342 & 20882   & -6.3828e-03 & -4.2552e-03 & 5  \\ 
	3 & 0.375 & 18788   & -7.0943e-03 & -4.7296e-03 & 5 \\ 
	4 & 0.3943 & 17713   & -7.5248e-03 & -5.0165e-03 & 5 \\ \hline     
   \textbf{5} & \textbf{0.3953} & \textbf{17660} & \textbf{-7.5474e-03} & \textbf{-5.0316e-03} & \textbf{5} \\ 
\textbf{5a} & \textbf{0.3953} & \textbf{17660}   & \textbf{-7.5474e-03} & \textbf{-5.0316e-03} & \textbf{10} \\ 
\textbf{5b} & \textbf{0.3953} & \textbf{17660}   & \textbf{-3.7737e-03} & \textbf{-2.5158e-03} & \textbf{5} \\ \hline
	6 & 0.3962 & 17612  & -7.5677e-03 & -5.0451e-03 & 5 \\     
	7 & 0.415 & 16618  & -7.9995e-03 & -5.3330e-03 & 5 \\     
    8 & 0.4384 & 15581  & -8.5544e-03 & -5.7029e-03 & 5 \\ 
    9 & 0.48 & 13890   & -9.5958e-03 & -6.3972e-03 & 5 \\
  \end{tabular}
  \caption{Parameter values in the eleven different numerical simulations we present for the uniform stratification. Modes 1 \& 2 at frequency $\omega_0$ are forced with amplitudes $W_1$ and $W_2$ (see equation \ref{scalings2}) at $x = 0$. $\lambda_1$ denotes the horizontal wavelength of mode-1 at frequency $\omega_0$. The computational grid for all the cases contained 4096 points in the horizontal to represent 5$\lambda_1$ (8192 points for case 5a) and 1024 points in the vertical to represent the ocean depth $H$.}
  \label{tab:Uniform stratification}
\end{table}

The primary wave frequencies corresponding to all the cases 1-9 are indicated in figure \ref{steady_uniform}(b), and are also shown in table \ref{tab:Uniform stratification}. The forcing amplitudes of modes 1 \& 2 at frequency $\omega_0$ (equations \ref{scalings1}-\ref{scalings3}) are also shown in table \ref{tab:Uniform stratification}. Except for case 5b, these amplitudes correspond to an overall primary wave energy flux of 2000W/m, with 90\% of it in mode-1. In case 5b, the forcing amplitudes for both modes 1 \& 2 are halved, thus corresponding to an overall primary wave energy flux of 500W/m. The horizontal and vertical spatial resolution in all the cases 1-9 are $dx\approx20$m and  $dz\approx3.7$m, respectively. These values correspond to around 820, 410 \& 270 grid points per wavelength of mode-$1$ at $\omega_0$, mode-$2$ at $\omega_0$ and mode-$1$ at $2\omega_0$, respectively. The temporal resolution of 2s in all the  cases 1-9 corresponds to around 725 time steps per time period of the superharmonic wave. The depth of the computational domain is fixed at $H = 3800$m, and the uniform stratification at $N_0 = 5.477 \times 10^{-3}$rad/s, both of which are somewhat representative of the ocean. The length of the computational domain is chosen to be around five times the horizontal wavelength of the primary wave mode-1 (except case 5a), which we find to be sufficiently long to capture the initial spatial growth of the superharmonic wave. The computational domain length is doubled for case 5a so as to demonstrate domain size independence of the initial growth rate of the superharmonic wave.  

\subsubsection{Nonuniform stratification}
\label{nu_cases}  

\begin{figure}
\centering
\includegraphics[width=1\textwidth]{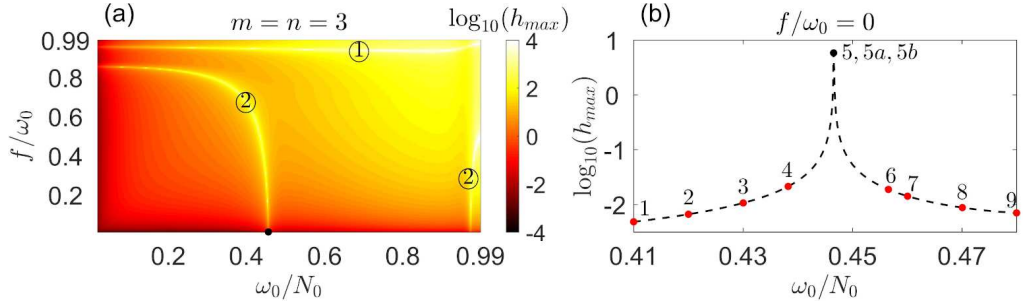}
\caption{(a) Steady-state amplitude of the superharmonic wave at 2$\omega_0$ (resulting from self-interaction of mode-3, see equation \ref{psi2_steady}) as a function of $\omega_0/N_0$ and $f/\omega_0$
in the nonuniform stratification $N_{max} = 4N_0$ (the encircled numbers denote the mode numbers of the superharmonic wave on the respective triadic resonance curves) , (b) Variation of $h_{max}$ (in log scale) with $\omega_0/N_0$ around the triadic resonance point (indicated by the solid black circle) for $f/\omega_0 = 0$. The red solid circles in (b) denote the off-resonance frequencies for which numerical simulations were performed, with the numbers next to them denoting the case numbers specified in table \ref{tab:nonuniform stratification}.}
\label{steady_nu}
\end{figure}

We consider the self-interaction of mode-3 at frequency $\omega_0$ in a nonuniform  stratification profile (ocean depth $H = 3800$m) given by
\begin{gather}
N(z) = N_0 + (N_{max}- N_0)\exp(-(z-z_c)^2/\sigma^2),
\label{strat}
\end{gather}
with the deep ocean stratification $N_0 = 5.477 \times 10^{-3}$rad/s, the maximum pycnocline stratification $N_{max} = 4N_0$ at $z_c = 3400$m, and the pycnocline half-width $\sigma = 250$m. As shown by \cite{JFM2017}, the self-interaction of mode-3 at frequency $\omega_0$ in the stratification given by equation (\ref{strat}) corresponds to three different curves on the $\omega_0/N_0-f/\omega_0$ plane on which triadic resonance occurs. A plot of the superharmonic wave amplitude from the steady-state theory (section \ref{offres}) in figure \ref{steady_nu}(a) shows these three different curves, where the mode number of the superharmonic internal wave is 1 or 2 (as indicated by the encircled numbers in figure \ref{steady_nu}a). In this paper, we perform numerical simulations at and around the solid black circle in figure \ref{steady_nu}(a), i.e. around one of the triadic resonance curves in the limit of no background rotation ($f=0$). Figure \ref{steady_nu}(b) shows the nine different frequencies at which numerical simulations are run, with cases 5, 5a and 5b being at the primary wave frequency for which triadic resonance occurs.   

The primary wave frequencies corresponding to all the cases 1-9 are shown in table \ref{tab:nonuniform stratification}. The forcing amplitudes of mode 3 at frequency $\omega_0$ (equations \ref{scalings1}-\ref{scalings3}) are also shown in table \ref{tab:nonuniform stratification}. Except for case 5a and 5b, these amplitudes correspond to an overall primary wave energy flux of 500W/m. In cases 5a and 5b (relatively longer domain for case 5b), the forcing amplitudes for mode 3 is made one-fifths, thus corresponding to an overall primary wave energy flux of 20W/m. The horizontal and vertical spatial resolution in all the cases 1-9 are  $dx\approx8$m and  $dz\approx3.7$m, respectively. These values correspond to around 820 \& 410 grid points per wavelength of mode-3 at $\omega_0$ and mode-2 at $2\omega_0$, respectively. The temporal resolution of 0.5s in all the cases 1-9 corresponds to around 2570 time steps per time period of the superharmonic wave. The length of the computational domain is chosen to be around five times the horizontal wavelength of the primary wave mode-3 (except case 5b), which we find to be sufficiently long to capture the initial spatial growth of the superharmonic wave. The computational domain length is doubled for case 5b so as to demonstrate domain size independence of the initial growth rate of the superharmonic wave.

\begin{table}
  \centering
  \renewcommand{\arraystretch}{1.25}
  \setlength{\tabcolsep}{4pt}
  \begin{tabular}{ c  c  c  c  c  c  c  c  c p{1cm} }
    Case & $\omega_0/N_0$  & $\lambda_1 = 2\pi/k_1(m)$  & $W_1(m/s)$ & $D/\lambda_1$  \\ \hline  
1 & 0.41 & 7617.8 & -9.2215e-03 & 5 \\
2 & 0.42 & 7409.3 & -9.4811e-03 & 5 \\
3 & 0.43 & 7209.7 & -9.7435e-03 & 5 \\    
4 & 0.4382 & 7052.5 & -9.9608e-03 & 5 \\     \hline 
\textbf{5} & \textbf{0.4466} & \textbf{6897.4} & \textbf{-10.0047e-03} & \textbf{5} \\ 
\textbf{5a} & \textbf{0.4466} & \textbf{6897.4} & \textbf{-2.037e-03} & \textbf{5} \\ 
\textbf{5b} & \textbf{0.4466} & \textbf{6897.4} & \textbf{-2.037e-03} & \textbf{10} \\\hline
6 & 0.4565 & 6720.43 & -10.4529e-03 & 5 \\ 
7 & 0.46 & 6659.7 & -10.5482e-03 & 5 \\ 
8 & 0.47 & 6490.9 & -10.8226e-03 & 5 \\ 
9 & 0.48 & 6328.5 & -11.1003e-03 & 5 \\ 
  \end{tabular}
  \caption{Parameter values in the eleven different numerical simulations we present for the nonuniform stratification ($N_{max}=4N_0$). Mode-3 at frequency $\omega_0$ is forced with amplitudes $W_1$ (see equation \ref{scalings2}) at $x = 0$. $\lambda_1$ denotes the horizontal wavelength of mode-3 at frequency $\omega_0$. The computational grid for all the cases contained 4096 points in the horizontal to represent 5$\lambda_1$ (8192 points for case 5c) and 1024 points in the vertical to represent the ocean depth $H$.}
  \label{tab:nonuniform stratification}
\end{table}

\subsection{Analysis of simulated wave fields}
\label{sec:numerical_analysis}
Temporally and spatially resolved horizontal and vertical velocity fields were written as outputs from the numerical simulation of each of the cases in tables \ref{tab:Uniform stratification} and \ref{tab:nonuniform stratification}. In the uniform stratification cases, the output was written at a temporal resolution of 8s, corresponding to around 360 points per one primary wave time period. For the nonuniform stratification cases, the temporal resolution was 16s, which corresponds to around 180 points per one primary wave time period. For both the uniform and nonuniform stratification cases, the vertical resolution of the written output was retained at the same value as that of  the computational grid, i.e. 1024 points to represent the entire ocean depth. In the horizontal, data was written at around 400 points that span the entire computational domain length. 

The uniform and nonuniform stratification simulations were typically run for a total time of around $40T$, which was verified to be long enough that the wave field amplitude at various frequencies and modes attains a steady state everywhere in the computational domain. For the longer domain cases (5a in table \ref{tab:Uniform stratification} and 5c in table \ref{tab:nonuniform stratification}) that were run to ensure domain size independence of our results, the simulation was run up to around 60$T$. At every spatial location, the time series of the velocity components from 34$T$ to 40$T$ were used to calculate the strength, and subsequently filter, at the primary (forcing) and superharmonic frequencies. 

The discrete Fourier transform of any flow quantity at a given spatial location is computed as
\begin{gather}
\widehat{W}(x,z,\Omega)  = \sum_{l=1}^{L} w(x,z,t_l)\exp(\frac{-2\pi i(k-1)(l-1)}{L}), \;\;\; 1\le k\le L,
\label{dft_defn}
\end{gather}
where $t_l$ denotes the uniformly distributed times at which the output data is written from the numerical simulations. The frequency $\Omega$ associated with index $k$ is given by $\Omega(k) = 2\pi(k-1)/(t_L-t_1)$ for $1\le k\le (1+L/2)$, with $\Omega(L-k+1)= \Omega(k+1)$ for $1\le k\le L/2$. The expression in equation (\ref{dft_defn}) is evaluated using the Matlab in-built function $fft$. The wave field filtered at $\Omega = \Omega_0$ is computed as
 \begin{gather}
w_{\Omega_0}(x,z,t_l)  = \frac{1}{L}\sum_{k=1}^{L} \widehat{W}_{\Omega_0}(x,z,\Omega(k))\exp(\frac{2\pi i(k-1)(l-1)}{L}), \;\;\; 1\le l\le L,
\label{idft_defn}
\end{gather}
where $\widehat{W}_{\Omega_0}(x,z,\Omega(k)) = \widehat{W}(x,z,\Omega(k))$ if $(\Omega_0-\Delta\Omega)\le\Omega(k)\le(\Omega_0+\Delta\Omega)$ and zero otherwise. For filtering at the forcing frequency $\Omega_0 = \omega_0$ and superharmonic frequency $\Omega_0 =2\omega_0$, we choose the half-width of the frequency window to be $\Delta\Omega  = 0.2\omega_0$, for all the uniform and nonuniform stratification cases. The expression in equation (\ref{idft_defn}) is evaluated using the Matlab in-built function $ifft$.

The filtered velocity fields (equation \ref{idft_defn}) are written as a superposition of left-to-right propagating modes at the corresponding frequency $\Omega_0$:
\begin{gather}
w_{\Omega_0}(x,z,t_l) = \sum_{j=1}^{M} W^{j}_{\Omega_0}\phi^{j}_{\Omega_0}(z)\sin(k^{j}_{\Omega_0}x-\Omega_{0}t_{l}+\gamma^{j}_{\Omega_0}),
\end{gather}
where $\phi^{j}_{\Omega_0}$ and $k^{j}_{\Omega_0}$ are the stream function vertical structure and the horizontal wavenumber associated with the $j^{th}$ mode at frequency $\Omega_0$. All the vertical mode shapes are normalized such that  they correspond to an energy flux of unity, as was done earlier in section \ref{sec:AEE}. The vertical velocity modal amplitudes $W^{j}_{\Omega_0}$ are related to the stream function amplitudes $\Psi_q$ in equation (\ref{psi1_sol}) via $|W^{j}_{\Omega_0}| = |k_q\Psi_q|$, with $j$ being the mode number associated with the $q^{th}$ wave in the resonant triad.

The output data from the numerical simulations are subjected to a modal decomposition analysis to estimate the horizontal evolution of the modal amplitudes, $W^{j}_{\Omega_0}(x)$. At a given $x  = x_0$, the filtered vertical velocity is represented as
\begin{gather}
w_{\Omega_0}(x_0,z,t) = A(z)\cos\Omega_{0}t + B(z)\sin\Omega_{0}t,
\end{gather} 
where $A(z)$ and $B(z)$ are chosen so as to minimize the error sum of squares of the fitted temporal variation at the discrete times spanning $t_1$ to $t_L$ at the spatial location $(x_0,z)$. The modal amplitude at $x=x_0$ is then estimated as
\begin{gather}
W^{j}_{\Omega_0}(x_0) = \frac{\sqrt{[\int_0^H(N^2-\Omega_0^2)A(z)\phi^{j}_{\Omega_0}(z)dz]^2 + [\int_0^H(N^2-\Omega_0^2)B(z)\phi^{j}_{\Omega_0}(z)dz]^2}}{\int_0^H(N^2-\Omega_0^2)(\phi^{j}_{\Omega_0}(z))^2dz},
\end{gather}  
where the orthogonality condition $\int_0^H(N^2-\Omega_0^2)(\phi^{j_1}_{\Omega_0})(\phi^{j_2}_{\Omega_0})dz = 0$ for $j_1\ne j_2$ has been used. A plot of $W^{p}_{2\omega_0}(x_0)$ with $x_0$ (recall from figure \ref{cartoon1} that $p$ is the mode number of superharmonic wave) allows us to estimate a slope, and hence validate the theoretical prediction for the early spatial growth of the superharmonic secondary wave while the primary wave amplitudes are assumed to be nearly constant. While the secondary superharmonic waves are not necessarily internal waves for $\omega_0$ away from the value at which triadic resonance occurs, we still perform the post-processing analysis described in this section for the off-resonance cases to quantify the behaviour away from triadic resonance. The analysis tools described in this section have also been implemented on the horizontal velocity field from each of our numerical simulations, and the corresponding results are entirely consistent with what we report in \S~\ref{sec:results} based on the vertical velocity fields. 

\section{Results}
\label{sec:results}
While the amplitude evolution equations derived in \S~\ref{sec:AEE} are valid for any resonant triad comprising three internal wave modes, we focus on the specific topic of spontaneous superharmonic generation due to triadic resonance between two primary internal wave modes at a frequency $\omega_0$ and a secondary superharmonic internal wave mode at frequency $2\omega_0$. The aim of this section is to study the dynamics of such resonant triads by analyzing the solutions of the amplitude evolution equations, and validate some of the theoretical results using direct numerical simulations. In \S~\ref{subsec:uniform}, we consider a scenario where two primary wave modes ($m=1,n=2$) at the same frequency in a uniform stratification form a resonant triad with a superharmonic internal wave mode ($p=1$). In \S~\ref{subsec:nonuniform}, we consider an idealized nonuniform stratification representative of the ocean, and investigate a scenario where a self-interacting primary mode-3 internal wave is in triadic resonance with a superharmonic mode-2 wave.

\subsection{Uniform stratification}
\label{subsec:uniform}

\begin{table}
  \centering
  \renewcommand{\arraystretch}{1.25}
  \setlength{\tabcolsep}{4pt}
\begin{tabular}{ c  c  c p{2cm} }
\bf{$\omega_1/N_0$} & \bf{$\omega_2/N_0$} & \bf{$\omega_3/N_0$} & \bf{$\alpha_1(sm^{-3})$}  \\
\bf{$k_1(m^{-1})$} & \bf{$k_2(m^{-1})$} & \bf{$k_3(m^{-1})$} & \bf{$\alpha_2(sm^{-3})$} \\
\bf{$m$} & \bf{$n$} & \bf{$p$}  & \bf{$\alpha_3(sm^{-3})$} \\
\hline
\bf{0.3953}  & \bf{0.3953}  & 0.7906 & 2.8079e-08 \\
\bf{3.558e-04} & \bf{7.116e-04} & 10.674e-04  & 2.8088e-08 \\
\bf{1} & \bf{2} & 1 & 5.6160e-08 \\
\end{tabular}
\caption{The frequencies ($\omega_j$), horizontal wavenumbers ($k_j$) and the mode numbers associated with the resonant triad in a uniform stratification considered in section \ref{subsec:uniform}. Specifically,  the subscripts 1 \& 2 denote the primary waves, whereas subscript 3 denotes the superharmonic secondary wave.}\label{T3}
\end{table}

In a uniform stratification ($N(z) = N_0$), mode-m and mode-n at frequency $\omega_0$ form a superharmonic resonant triad with mode-p ($=|m-n|$) at frequency $2\omega_0$ if the following condition is satisfied \citep{JFM2017}:
\begin{gather}
\frac{\omega_0^2}{N_0^2} = \frac{(m+n)^2 - (m-n)^2(4 - f^2/\omega_0^2)/(1-f^2/\omega_0^2)}{4(m+n)^2 - (m-n)^2(4 - f^2/\omega_0^2)/(1-f^2/\omega_0^2)}. \label{diverge_condn}
\end{gather}
As mentioned in \S~\ref{sec:theory}, the primary wave field (modes $m$ \& $n$ at frequency $\omega_0$) is assumed to derive its finite initial energy from physical mechanisms such as internal tide generation or scattering by ocean floor topography. Specifically, we assume that the total energy flux $E_T$ in the primary wave
field at frequency $\omega_0$ is divided as $(1-\beta)E_T$ in mode-$m$ and $\beta E_T$ in mode-$n$. Furthermore, the superharmonic secondary wave is assumed to have zero initial energy flux in the theoretical calculations, and no explicit noise is added in the numerical simulations. In this subsection, we focus on the specific case of $m = 1$, $n = 2$ and $f = 0$. Substituting $m = 1$, $n = 2$ and $f = 0$ in equation (\ref{diverge_condn}), we obtain $\omega_0/N_0 = 0.3953$. In the notation of section \S~\ref{sec:AEE}, the corresponding resonant triad is characterized by the values indicated in the first three columns of table \ref{T3}.

\subsubsection{Theory}
\label{ut_results}
For a uniform stratification, the coefficients in the amplitude evolution equations (\ref{amp1_eqn})-(\ref{amp3_eqn}) for a resonant triad can be evaluated analytically; for the sake of brevity, we do not show the final analytical expressions here. We proceed to calculate the specific values of these coefficients using representative values of $H = 3800$m, $N_0 = 5.477 \times 10^{-3}$rad/s, $\bar{\rho} = 1000$kg/m$^3$ and the normalised mode shapes corresponding to energy flux in each wave mode being unity are given by
\begin{gather}
\phi_j(z) = \sqrt{\frac{4\omega_j}{\bar{\rho}H(N_0^{2}-\omega^{2}_j)k_j}}\sin(k_j\cot\theta_jz).
\label{phi_uniform}
\end{gather}
The values of the coefficients $\alpha_1$, $\alpha_2$ and $\alpha_3$  for the resonant triad considered in this section are indicated in the fourth column of table \ref{T3}. 

For a given resonant triad, the analytical solution of the amplitude evolution equations for small $X$ can be obtained by assuming that the primary wave amplitudes remain spatially invariant. This assumption is reasonable since the initial amplitudes of the primary waves are typically much larger than those for the secondary waves. Specifically, the initial spatial evolution of the secondary wave amplitude in the resonant triad is given by
\begin{gather}
\Psi_{3}(X) = \Psi_{3}(X=0) + i\alpha_{3} \Psi_{1}(X=0) \Psi_{2}(X=0) X,
\label{psi3_sup}
\end{gather}
where $\Psi_{1}(X=0) = \sqrt{(1-\beta)E_T}$ , $\Psi_{2}(X=0) = \sqrt{\beta E_T}$ and $\Psi_{3}(X=0) = 0$. The rate of initial growth of $|\Psi_3|$ with $X$ is hence given by $\Gamma = \alpha_3E_T\sqrt{\beta(1-\beta)}$. Assuming the initial growth rate $\Gamma$ to sustain for large distances, the quantity $|\Psi_3|^2$ (the energy flux in the superharmonic wave) would grow to a fraction $\delta$ of the initial primary wave energy flux $E_T$ at
\begin{gather}
X_{\delta} = \sqrt{\frac{\delta}{\alpha^{2}_{3}\beta(1-\beta)E_T}}.
\end{gather}
For $E_T = 2000$W/m, $\beta = 0.1$, $X_\delta/\lambda_1$ $\approx$ 24, 53 and 75 for $\delta=$ 0.1, 0.5 and 1, respectively ($\lambda_1$ is the horizontal wavelength of the primary mode-1 wave). Increasing $\beta$ to 0.5 with $E_T$ remaining at 2000W/m, the corresponding values of $X_\delta/\lambda_1$ decrease to $\approx$ 14, 32 and 45. While  the initial growth rate estimates suggest that the growth of the superharmonic wave can be significant for values of $E_T$ and $\beta$ representative of the ocean, the assumption of sustained growth at the initial growth rate $\Gamma$ is invalid for large X. We therefore proceed to plot the numerical solutions of the amplitude evolution equations where $\Psi_1$, $\Psi_2$ and $\Psi_3$ simultaneously evolve.

We normalize the various quantities in the amplitude evolution equations (\ref{amp1_eqn})-(\ref{amp3_eqn}) to re-write them as
\begin{gather}
\frac{d\widetilde{\Psi}_1}{dX} = i\widetilde{\alpha}_1\widetilde{\Psi}^{*}_2\widetilde{\Psi}_3,\label{amp1_tilda}\\
\frac{d\widetilde{\Psi}_2}{dX} = i\widetilde{\alpha}_2\widetilde{\Psi}^{*}_1\widetilde{\Psi}_3,\label{amp2_tilda}\\
\frac{d\widetilde{\Psi}_3}{dX} = i\widetilde{\alpha}_3\widetilde{\Psi}_1\widetilde{\Psi}_2,
\label{amp3_tilda}
\end{gather}
where $(\widetilde{\Psi}_1,\widetilde{\Psi}_2,\widetilde{\Psi}_3) = (\Psi_1,\Psi_2,\Psi_3)/\sqrt{E_T}$ and $(\widetilde{\alpha}_1,\widetilde{\alpha}_2,\widetilde{\alpha}_3) = \sqrt{E_T}(\alpha_1,\alpha_2,\alpha_3)$. The initial conditions, written in terms of the normalized quantities, are $\widetilde{\Psi}_1(X =0) = \sqrt{1-\beta}$, $\widetilde{\Psi}_2(X =0) = \sqrt{\beta}$ and $\widetilde{\Psi}_3(X =0) = 0$. 

\begin{figure}
\centering
\includegraphics[width=1\textwidth]{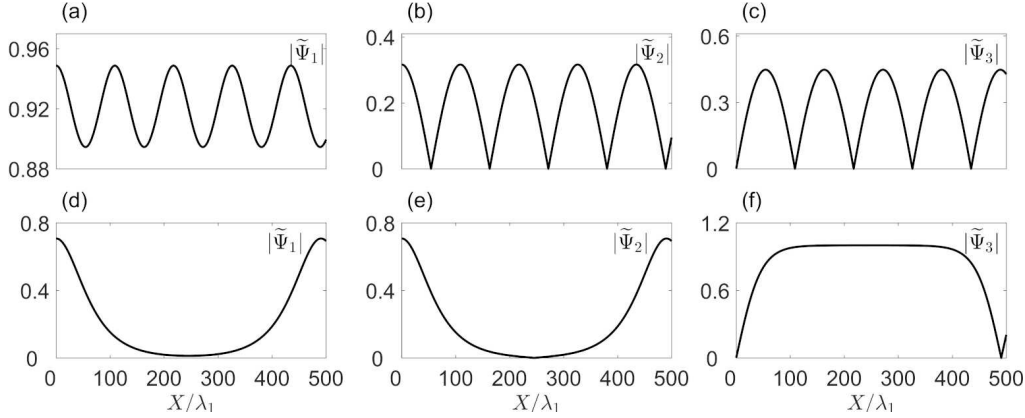}
\caption{Numerical solutions of the amplitude evolution equations (\ref{amp1_tilda})-(\ref{amp3_tilda}) for the resonant triad in a uniform stratification considered in section \ref{subsec:uniform}. The spatial evolution of (a) $|\widetilde{\Psi}_1|$, (b) $|\widetilde{\Psi}_2|$ and (c) $|\widetilde{\Psi}_3|$ for $E_T = 2000$W/m, $\beta = 0.1$ (d)-(f) are the same as (a)-(c) but for $\beta = 0.5$. }\label{fig_ut1}
\end{figure}

Figure \ref{fig_ut1} shows the numerical solutions of equations (\ref{amp1_tilda})-(\ref{amp3_tilda}) for the resonant triad listed in table \ref{T3}, and with $E_T = 2000$W/m. For $\beta = 0.1$ (top row of figure \ref{fig_ut1}), the mode-2 primary wave amplitude decays to zero at $X/\lambda_1 = 54$ (figure \ref{fig_ut1}b), while the mode-1 superharmonic wave amplitude reaches its maximum value of $0.447$ (figure \ref{fig_ut1}c). Over the same horizontal distance, the mode-1 primary wave loses around $11.12$\% of its energy (figure \ref{fig_ut1}a). After reaching their respective minimum values, the primary waves start to increase in their amplitude towards their initial values while the superharmonic wave amplitude decreases towards 0. The initial wave field is then seen again at $X/\lambda_1 = 108$. This periodic behaviour then persists for larger $X$ too.

In the bottom row of figure \ref{fig_ut1}, we plot the theoretical amplitude evolution for $\beta = 0.5$. With $\tilde{\Psi}_1 = \tilde{\Psi}_2$ at $X = 0$, and $\alpha_1\approx\alpha_2$ (see table \ref{T3}), the evolution of $\tilde{\Psi}_1$ and $\tilde{\Psi}_2$ are quite similar. Both $\tilde{\Psi}_1$ and $\tilde{\Psi}_2$ decrease to around $21.4$ \% of their initial value over around $X_1/\lambda_1 = 100$, and remain even smaller for a further distance of around 290$\lambda_1$. While the decay distance for $\beta = 0.5$ is larger than that for $\beta = 0.1$, it is noteworthy that the superharmonic wave for $\beta = 0.5$ possesses almost the entire primary wave energy over significantly large distances.

\begin{figure}
\centering
\includegraphics[width=1\textwidth]{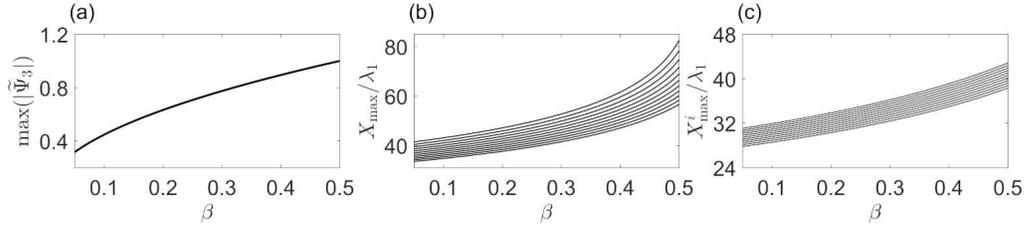}
\caption{(a) The maximum amplitude of the superharmonic wave $|\widetilde{\Psi}_3|$ obtained from the numerical solution of equations (\ref{amp1_tilda})-(\ref{amp3_tilda}) for various $\beta$. The square of the plotted quantity in (a) is the energy flux in the superharmonic wave (as a fraction of the total energy $E_T$). (b) The horizontal distance at which $|\widetilde{\Psi}_3|$ reaches $\Delta\times \max(|\tilde{\Psi}_3|)$. $\Delta$ varies uniformly from 0.85 to 0.95 from the bottom to the top curves. (c)  Same as in (b), if $|\widetilde\Psi_3|$ were to continuously grow with its initial growth rate $\Gamma$. Figures (b) and (c) correspond to $E_T = 2000$W/m, whereas figure (a) is independent of $E_T$. }\label{fig_ut2}
\end{figure}

From the numerical solutions such as those in figure \ref{fig_ut1}, we estimate the maximum value that $|\widetilde{\Psi}_3|$ reaches, and plot it as a function of $\beta$ in figure \ref{fig_ut2}(a). It is noteworthy that figure \ref{fig_ut2}(a) is independent of $E_T$, and the square of the plotted quantity is the maximum fraction of the initial primary wave energy that gets transferred to the superharmonic wave. While around 20\% of the initial primary wave energy gets transferred to the superharmonic wave for $\beta = 0.1$, it becomes a substantial 60\% for $\beta = 0.3$. Finally, for $\beta = 0.5$, all the primary wave energy gets transferred to the superharmonic wave, as was already seen in the bottom row of figure \ref{fig_ut1}. The horizontal distance, $X_{max}$ at which the superharmonic wave amplitude reaches $\Delta$ of its maximum value is plotted as a function of $\beta$ in figure \ref{fig_ut2}(b). $\Delta$ varies uniformly from 0.85 to 0.95 from the bottom to the top curves in figure \ref{fig_ut2}(b). Figure \ref{fig_ut2}(b) indicates that a substantial amount of the initial primary wave energy is transferred to the superharmonic wave by around 50$\lambda_1$. In the fictitious limit of sustained growth of the superharmonic wave at its initial growth rate $\Gamma$, the distances plotted in figure \ref{fig_ut2}(b) would be smaller, as shown in figure \ref{fig_ut2}(c). Finally, we note that the horizontal location where $|\widetilde\Psi_3|$ grows to its maximum amplitude scales as $1/\sqrt{E_T}$, giving rise to slower superharmonic wave growth for smaller $E_T$.   

As an overall summary, this subsection has shown that a significant fraction of the primary wave energy flux could get transferred to the secondary superharmonic wave, over distances that are a few tens of the primary mode-1 wavelength if we consider parameter values that are somewhat representative of the ocean. This superharmonic wave generation mechanism should therefore be a consideration along with other triadic resonances like the parameteric subharmonic instability, a topic we discuss further in \S~\ref{sec:disc}. We now proceed to validate our theoretical results, i.e. the generation of a superharmonic wave and its corresponding spatial growth, using direct numerical simulations. In the numerical simulations, we simulate only the initial growth of the superharmonic wave, and leave the topic of the wave field evolution over very large distances to a future study.             

\subsubsection{Simulations}
\label{u_results}

The details of the numerical simulations are described in \S~\ref{sec:numerics}, and the specific cases we run for a uniform stratification are listed in table \ref{tab:Uniform stratification}. While cases 5, 5a and 5b correspond exactly to the primary wave frequency at which triadic resonance occurs, the other cases correspond to off-resonance scenarios. Figure \ref{steady_uniform}(b) shows the relative positioning of the various off-resonance cases with respect to the primary wave frequency at which triadic resonance occurs, i.e. cases 5, 5a and 5b.

\begin{figure}
\centering
\includegraphics[width=1\textwidth]{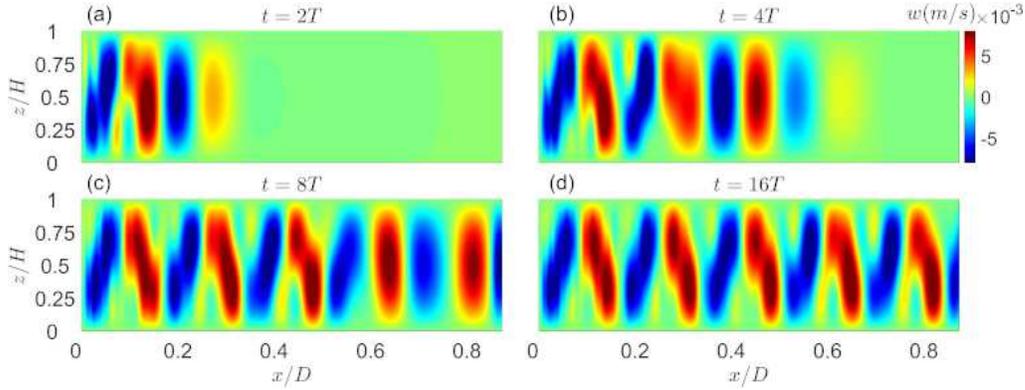}
\caption{Instantaneous vertical velocity field at (a) $t = 2T$, (b) $t = 4T$, (c) $t = 8T$ and (d) $t = 16T$ from the numerical simulation of case-5 in table I for the uniform stratification. $T  = 2\pi/\omega_0$ is the time period of the modes 1 \& 2 forced at $x = 0$. Steady state characteristics of the wave field are computed from the simulation data during $t = 34T$ to $t = 40T$.}
\label{vel_inst}
\end{figure}

Figure \ref{vel_inst} shows the evolution of the numerically simulated vertical velocity field for case-5, i.e. the triadic resonance case. Modes 1 \& 2 at the primary wave frequency $\omega_0$ are forced at $x = 0$, and it is evident from figures \ref{vel_inst}(a)-(c) that mode-1 propagates faster than mode-2. At $t = 16T$, both the primary wave modes seem to have reached the right end of the computational domain, where a sponge layer absorbs all the incoming wave energy (figure \ref{vel_inst}d). The simulations were run for a further time of 24$T$ to ensure that steady state is reached throughout the domain with respect to the primary and secondary wave amplitudes. Finally, the simulated flow from 34T-40T is considered for further analysis of the steady state wave field.   

\begin{figure}
\centering
\includegraphics[width=1\textwidth]{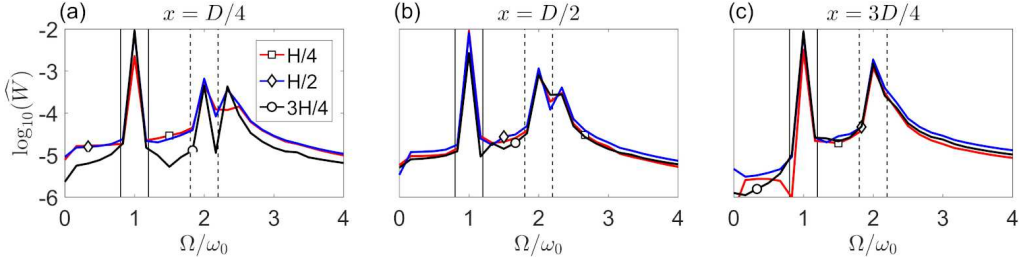}
\caption{{\it Fourier spectra of the vertical velocity at specific spatial locations from the numerical simulation of case-5 in table \ref{tab:Uniform stratification} for the uniform stratification.} FFT ($\widehat{W}$) of  the vertical velocity time series from $t = 34T$ to $t = 40T$ at (a) x = D/4, (b) x = D/2 and (c) x = 3D/4 . In each of the plots, the three curves correspond to $z = H/4, H/2$ and $3H/4$. The solid and dashed vertical lines denote the frequency bands that are used to filter the wave field at $\Omega = \omega_0$ and $\Omega = 2\omega_0$, respectively.} \label{un_fft}
\end{figure}

Figure \ref{un_fft} shows the steady state Fourier spectrum (see equation (\ref{dft_defn}) for the definition of $\widehat{W}$) at nine different locations in the computational domain. A finite energy at the superharmonic frequency of $\Omega = 2\omega_0$ is evident at all locations, with the corresponding strength of $\widehat{W}$ growing from $x = D/4$ (figure \ref{un_fft}a) to $x = 3D/4$ (figure \ref{un_fft}c). Additionally, the region close to $x = 0$ seems to contain some energy away from $\Omega = 2\omega_0$ as well, specifically $\Omega\approx 2.35\omega_0$ at $x = D/4$, which we attribute to nonlinearities that may be associated with our forcing mechanism at x = 0. While figure \ref{un_fft} confirms the generation of the superharmonic frequency, it is important to calculate the spatial structure at $\Omega = 2\omega_0$ to  verify if it is indeed a superharmonic mode-1 internal wave as predicted by theory. We therefore filter the numerically simulated wave fields at $\Omega = \omega_0$ and $2\omega_0$, with the corresponding widths of the frequency bands indicated by the solid and dashed vertical lines, respectively, in figure \ref{un_fft}.   

\begin{figure}
\centering
\includegraphics[width=1\textwidth]{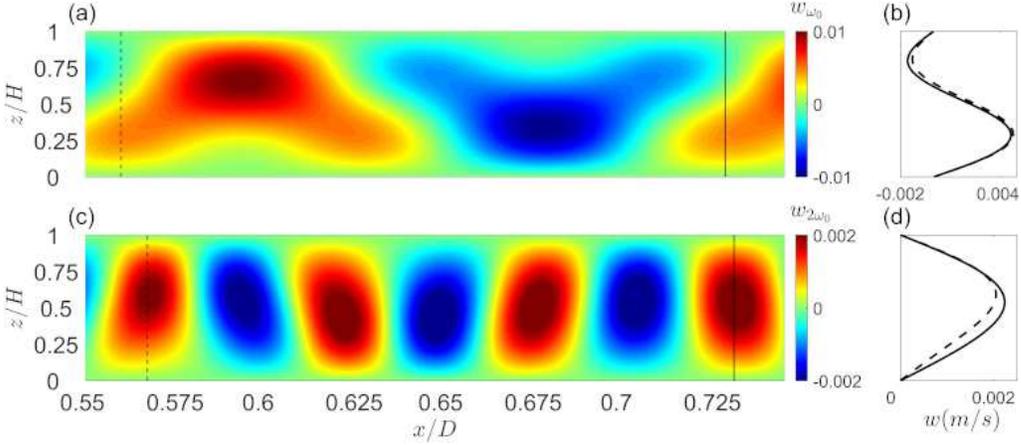}
\caption{{\it Filtered vertical velocity fields from the numerical simulation of case-5 in table \ref{tab:Uniform stratification} for the uniform stratification.} (a) Vertical velocity field $w_{\omega_0}$ at $t = 34T$, after filtering the raw time series during $t = 34T-40T$ at $\Omega = \omega_0$. The bandwidth of frequency used for the filtering is indicated by the vertical solid lines in figure \ref{un_fft}. (b) Vertical profiles of $w_{\omega_0}$ at the two spatial locations corresponding to the vertical dashed and solid lines in (a).  (c) Vertical velocity field $w_{2\omega_0}$ at $t = 34T$, after filtering the raw time series during $t = 34T-40T$ at $\Omega = 2\omega_0$. The bandwidth of frequency used for the filtering is indicated by the vertical dashed lines in figure \ref{un_fft}. (d) Vertical profiles of $w_{2\omega_0}$ at the two spatial locations corresponding to the vertical dashed and solid lines in (c).}\label{un_filt}
\end{figure}

\begin{figure}
\centering
\includegraphics[width=1\textwidth]{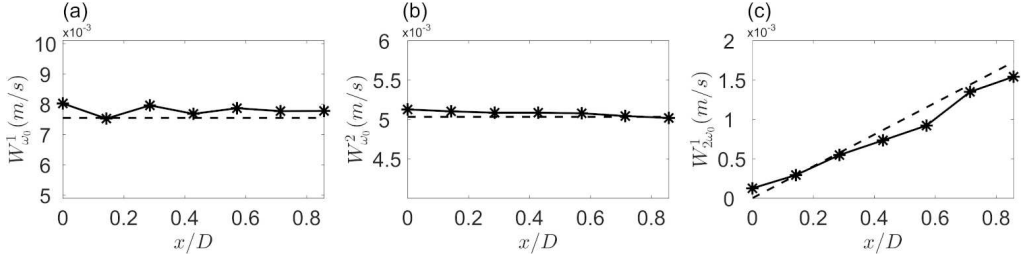}
\caption{Variation with $x$ of (a) mode-1, (b) mode-2 amplitudes at $\Omega = \omega_0$ based on the filtered vertical velocity field shown in figure \ref{un_filt}(a) from the numerical simulation of case-5 in table \ref{tab:Uniform stratification} for the uniform stratification. The dashed lines in (a) \& (b) indicate the corresponding forcing amplitudes at $x=0$. (c) Variation with $x$ of mode-1 amplitude at $\Omega = 2\omega_0$ based on the filtered vertical velocity field shown in figure \ref{un_filt}(c). The dashed line in (c) represents the linear growth of $W^{1}_{2\omega_0}$ as predicted by the amplitude evolution equations assuming the primary wave amplitudes $W^{1}_{\omega_0}$ and $W^{2}_{\omega_0}$ to be constant at the values shown in (a) \& (b). }\label{us_2w}
\end{figure}

Figure \ref{un_filt}(a) shows the vertical velocity field filtered at the primary wave frequency $\Omega = \omega_0$, with the filtering process described in detail in \S~\ref{sec:numerical_analysis}. The spatial structure of $w_{\omega_0}$ in figure \ref{un_filt}(a) corresponds to a combination of modes 1 \& 2, consistent with the forcing provided at $x = 0$. The vertical profiles of $w_{\omega_0}$ at two $x$ locations separated by one mode-1 wavelength show a combination of mode-1 \& mode-2, and are almost identical (figure \ref{un_filt}b). The spatial structure of the wave field filtered at the superharmonic frequency ($2\omega_0$) clearly shows a predominant mode-1, and the wavelength of $\approx 6km$ in $w_{2\omega_0}$ is fully consistent with the theoretically estimated horizontal wavelength of mode-1 at frequency $2\omega_0$ (figure \ref{un_filt}c). Figure \ref{un_filt}(c) thus provides numerical evidence of the generation of mode-1 superharmonic internal wave as a result of resonant interaction between modes 1 \& 2 at the primary wave frequency.

In figure \ref{us_2w}, we plot the vertical velocity amplitude of the primary modes 1 \& 2, and the superharmonic mode-1 as a function of $x$. The definition of the modal amplitudes at $\omega_0$ and 2$\omega_0$, and the methodology used to compute them from numerical simulations are discussed in \S~\ref{sec:numerical_analysis}. Figures \ref{us_2w}(a) and (b) show that the amplitude of primary modes 1 \& 2 remain almost constant over the entire domain, and their values are consistent with the amplitudes forced at $x = 0$. Small spatial variations in the primary modal amplitudes suggest that the superharmonic wave would grow with the initial growth rate $\Gamma$ (see \S~\ref{ut_results}) throughout the domain. Indeed, figure \ref{us_2w}(c) confirms that the vertical velocity amplitude of mode-1 at frequency $2\omega_0$ increases almost linearly with $x$. Furthermore, the theoretical initial growth rate of the superharmonic wave based on the amplitude evolution equations captures the simulated growth rate quite accurately. In summary, figure \ref{us_2w} provides quantitative validation of the initial amplitude evolution predicted by the theory presented in \S~\ref{ut_results}. As shown in appendix \ref{convergence}, numerical simulation of the resonant case 5 with a longer domain, i.e. case 5a in table \ref{tab:Uniform stratification}, reproduces the same initial growth rate for the superharmonic wave. Finally, numerical simulation of the resonant case 5 with smaller forcing amplitudes, i.e. case 5b in table \ref{tab:Uniform stratification}, also show an initial superharmonic wave growth that quantitatively agrees with the prediction of the amplitude evolution equations (appendix \ref{convergence}).  

\begin{figure}
\centering
\includegraphics[width=1\textwidth]{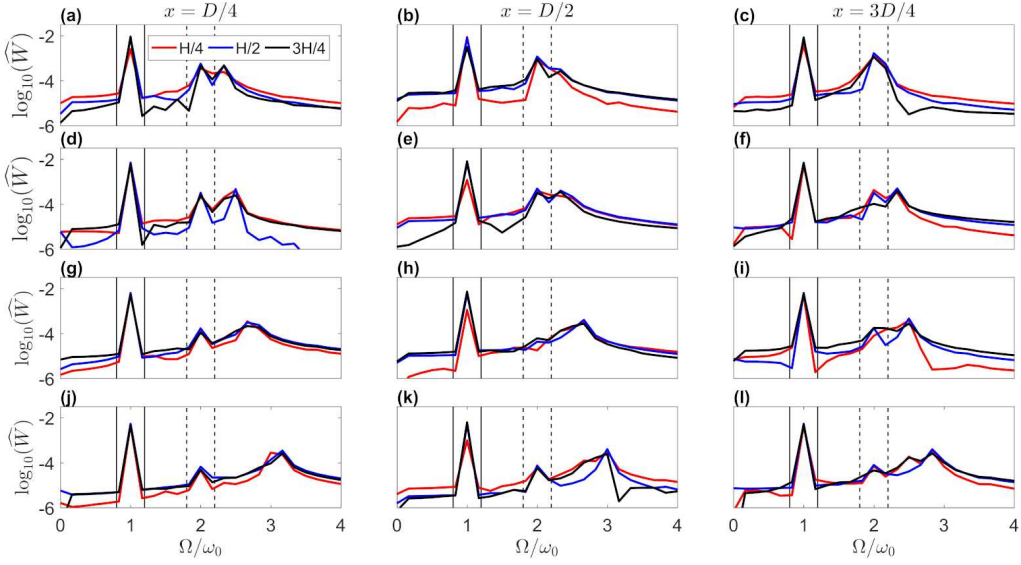}
\caption{{\it Fourier spectra of the vertical velocity at specific spatial locations from the numerical simulations of the off-resonance cases: case-4 (top row),  case-3 (second row), case-2 (thrid row) and case-1 (bottom row) in table \ref{tab:Uniform stratification} for the uniform stratification}. FFT($\widehat{W}$) of the vertical velocity time series from t = 34T to t = 40T at x = D/4 (left column), x = D/2 (middle column) and x = 3D/4 (right column). In each of the plots, the three curves correspond to z = H/4,H/2 and 3H/4. The solid and dashed vertical lines denote the frequency bands that are used to filter the wave field at  $\Omega = \omega_0$ and $\Omega = 2\omega_0$, respectively.}\label{u_offres}
\end{figure}

We proceed to numerically investigate the off-resonance cases listed in table \ref{tab:Uniform stratification}, i.e. cases 1-4 and 6-9. It is worthwhile recalling that the amplitude evolution equations are valid only at the resonant frequency, and hence cannot be used to predict the results from the off-resonance cases. Figure \ref{u_offres} shows the steady state Fourier spectra of the vertical velocity field for cases 1-4, with the topmost row being closest to the resonant case 5. The generation of the superharmonic frequency $2\omega_0$ is evident in case-4, with the corresponding amplitude being comparable to that of case-5. Superharmonic wave generation is observed in cases 1-3 too, with the corresponding amplitudes becoming weaker as we move further away from the resonant case-5. Interestingly, the off-resonance cases 1 and 2 show the presence of frequencies larger than $2\omega_0$, the origins of which are not clear. To quantify the generation of superharmonic ($2\omega_0$) waves in the off-resonance cases, we perform the modal decomposition analysis described in \S~\ref{methodology}, though the off-resonance superharmonic waves are not internal waves.

\begin{figure}
\centering
\includegraphics[width=1\textwidth]{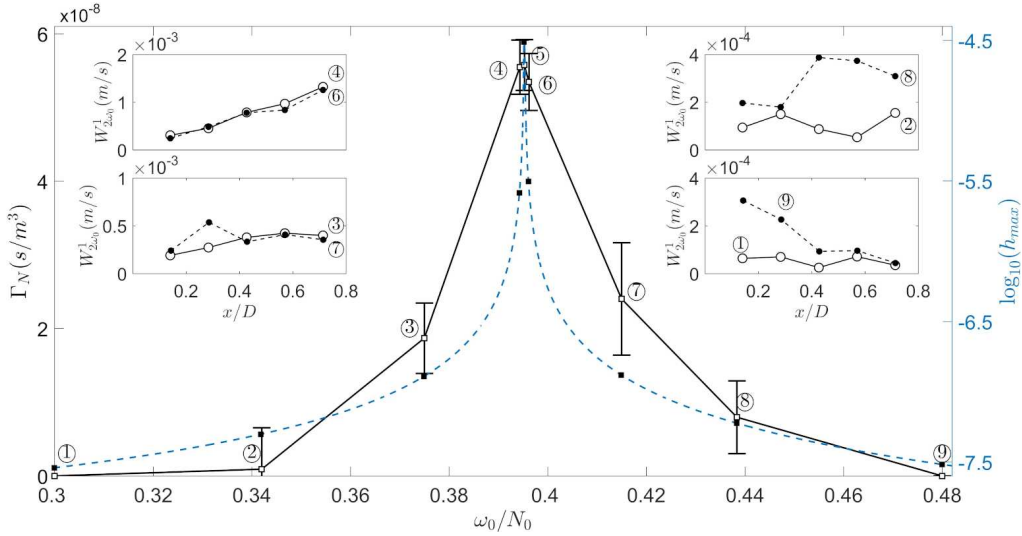}
\caption{The normalized growth rate ($\Gamma_N = \Gamma/(|\Psi_1(X=0)||\Psi_2(X=0)|$), see section \ref{ut_results} for the definition of $\Gamma$) of mode-1 at $\Omega = 2\omega_0$, estimated from the variation of $W_{2\omega_0}^1$ with $x$, plotted as a function of $\omega_0/N_0$ based on numerical simulations of all the cases in table \ref{tab:Uniform stratification} for the uniform stratification. The encircled numbers next to every data point represent the corresponding case numbers. The insets show the $W_{2\omega_0}^1$ vs. $x/D$ plots for cases 1-9, except the resonant case-5 whose variation is shown in figure 9. The blue dashed curve (with the black solid markers) shows the variation of the amplitude of the superharmonic mode-1 ($log_{10}(h_{max})$, plotted on the right axis) based on the steady-state theory presented in section \S~\ref{offres}.}\label{u_grand}
\end{figure}

Figure \ref{u_grand} shows the variation of the computed initial growth rate $\Gamma$ of the mode-1 superharmonic wave with the primary wave frequency $\omega_0$. The insets in figure \ref{u_grand} show the spatial variation of $W_{2\omega_0}^1$ for all the off-resonance cases. For cases 4 and 6, which are closest to the resonant case-5, the superharmonic wave growth is almost as strong as the resonant case-5. Moving further away from resonance, the superharmonic wave growth decreases, before becoming negligible for cases 1 and 9. Interestingly, the width of the peak in $\Gamma$ seems larger than the width of the peak in the superharmonic wave amplitude based on the steady state theory in \S~\ref{offres}. Figure \ref{u_grand} thus shows that the resonant generation of superharmonic internal waves at $\omega_0/N_0 = 0.3953$ has its signature even at off-resonance frequencies that are around $0.04N_0$ away from the resonant frequency.  

\subsection{Nonuniform Stratification}
\label{subsec:nonuniform}

In a typical nonuniform stratification representative of the ocean, \cite{JFM2017} showed that a larger number of triadic resonances resulting from modal interactions at a fixed frequency are possible when compared to a uniform stratification. Under certain conditions, \cite{JFM2017} showed that even an isolated internal wave mode (of arbitrary amplitude) may be inherently unstable; such a triadic resonance resulting from self-interaction of an isolated mode is not possible in a uniform stratification. In this section, we investigate the amplitude evolution associated with one such self-interaction scenario.

In the nonuniform stratification given by equation (\ref{strat}), \cite{JFM2017} showed that a self-interacting mode-3 at $\omega/N_0 = 0.4466$ is in triadic resonance with mode-2 at the superharmonic frequency $2\omega$ for $f = 0$. We solve the amplitude evolution equations corresponding to this case, and also perform DNS at and around the resonant frequency. 

\subsubsection{Theory}
\label{nut_results}
The frequencies and horizontal wavenumbers associated with the three waves that make up the resonant triad are listed in table \ref{T4}. As mentioned earlier, owing to the self-interaction case being considered, waves 1 and 2 both represent mode-3 at $\omega/N_0 = 0.4466$. The values of the coefficients $\alpha_1$, $\alpha_2$ and $\alpha_3$ (equations \ref{alpha1}-\ref{alpha3})
are also listed in table \ref{T4}, with $\alpha_3$ being twice of what is given by equation (\ref{alpha3}). Since waves 1 and 2 are the same, $4|\Psi_1|^2$ represents the energy flux in the primary mode-3 at $\omega$ while $|\Psi_3|^2$ represents the energy flux in the superharmonic mode-2.

Assuming $\Psi_1$ to remain constant for small $X$, the early spatial evolution of $\Psi_3$ can be written as
\begin{equation}
\Psi_3(X) = \Psi_3(X = 0) + i\alpha_3[\Psi_1(X=0)]^2 X,
\end{equation}
where $\Psi_1(X=0) = \sqrt{E_T}/2$ and $\Psi_3(X=0) = 0$. Using the value of $\alpha_3$ listed in table \ref{T4}, and $E_T = 2000$W/m, this initial growth would transfer around 17 \%, 30 \% and 47 \% of the initial energy to the superharmonic mode-2 over 30, 40 and 50 wavelengths of the primary mode-3, respectively if the initial growth rate sustains. The primary wave amplitude, however, does not remain constant, and we therefore compute the numerical solution of the amplitude evolution equations to estimate the growth of the superharmonic wave over large horizontal distances.

\begin{table}
  \centering
  \renewcommand{\arraystretch}{1.25}
  \setlength{\tabcolsep}{4pt}
\begin{tabular}{ c  c  c p{2cm} }
\bf{$\omega_1/N_0$} & \bf{$\omega_2/N_0$} & \bf{$\omega_3/N_0$} & \bf{$\alpha_1(sm^{-3})$}  \\
\bf{$k_1(m^{-1})$} & \bf{$k_2(m^{-1})$} & \bf{$k_3(m^{-1})$} & \bf{$\alpha_2(sm^{-3})$} \\
\bf{$m$} & \bf{$n$} & \bf{$p$}  & \bf{$\alpha_3(sm^{-3})$} \\
\hline
\bf{0.4466}  & \bf{0.4466}  & 0.8932 & 4.4529e-08 \\
\bf{9.1024e-04} & \bf{9.1024e-04} & 18.2152e-04  & 4.4529e-08 \\
\bf{3} & \bf{3} & 2 & 17.7587e-08 \\
\end{tabular}
\caption{The frequencies ($\omega_j$), horizontal wavenumbers ($k_j$) and the mode numbers associated with the resonant triad in a nonuniform stratification considered in section \ref{subsec:nonuniform}. Specifically,  the subscripts 1 \& 2 denote the primary waves, whereas subscript 3 denotes the superharmonic secondary wave.}\label{T4}
\end{table}

\begin{figure}
\centering
\includegraphics[width=1\textwidth]{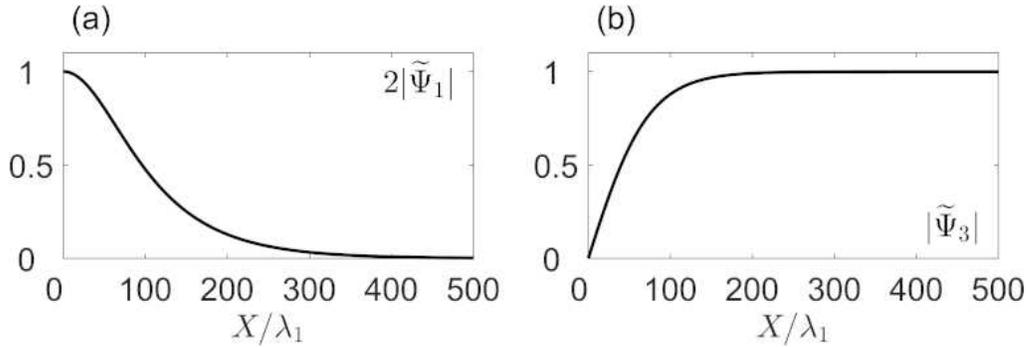}
\caption{Numerical solutions of the amplitude evolution equations (\ref{amp1_tilda})-(\ref{amp3_tilda}) for the resonant triad in a nonuniform stratification considered in section \ref{subsec:nonuniform}. The spatial evolution of (a) 2$|\widetilde{\Psi}_1|$ and (b) $|\widetilde{\Psi}_3|$ for $E_T = 2000$W/m.}\label{theory_nu}
\end{figure}

As done in section \ref{subsec:uniform}, we normalize $\Psi_1$, $\Psi_2$ and $\Psi_3$ by $\sqrt{E_T}$ so that $4|\widetilde{\Psi}_1|^2$ and $|\widetilde{\Psi}_3|^2$ represent the fraction of the initial primary wave energy flux $E_T$ that is in the primary and secondary (superharmonic) waves, respectively. Figure \ref{theory_nu} shows the spatial evolution of 2$|\tilde{\Psi}_1|$ and  $|\tilde{\Psi}_3|$, as obtained by numerically solving the amplitude evolution equations (\ref{amp1_eqn})-(\ref{amp2_eqn}) for $E_T = 2000$W/m. Over 30, 40 and 50 wavelengths of the primary mode-3, around 15 \%,  25\% and  35\% of the overall energy is transferred to the superharmonic mode-2. Remarkably, over a sufficiently large distance, almost all the primary wave energy is transferred to the superharmonic wave, and the wave field switches permanently from the state of all the energy being in the primary mode-3 to the state of all the energy being in the secondary superharmonic wave. In other words, the primary mode-3 at frequency $\omega/N_0 = 0.4466$ is inherently unstable, and would spontaneously transfer all its energy to the superharmonic mode-2. Similar to our conclusion in \S~\ref{ut_results}, for other values of $E_T$, qualitatively similar dynamics occur, but over spatial extents that are inversely proportional to $\sqrt{E_T}$.   

\subsubsection{Simulations}
\label{nu_results}

Direct numerical simulations are performed with two different aims: (i) to validate the theoretically estimated early spatial growth of the superharmonic wave in \S~\ref{nut_results}, and (ii) to investigate the wave field evolution at off-resonant frequencies, i.e. frequencies away from $\omega/N_0 = 0.4466$. The details of all the cases run for the nonuniform stratification are listed in Table \ref{tab:nonuniform stratification}, with case 5 corresponding to the frequency at which triadic resonance occurs. As mentioned in section \ref{nu_cases}, the forcing amplitudes for the nonuniform stratification cases were such that the total energy flux was 500W/m, which allowed us to accurately capture the weakly nonlinear regime with the computational grid size and time step values given in \S~\ref{nu_cases}. 

\begin{figure}
\centering
\includegraphics[width=1\textwidth]{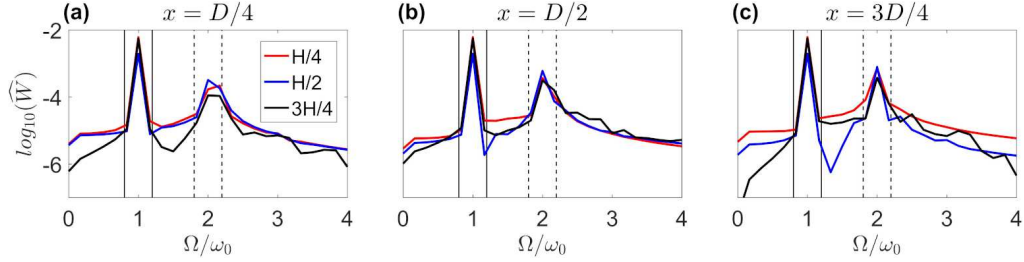}
\caption{{\it Fourier spectra of the vertical velocity at specific spatial locations from the numerical simulation of case-5 in table \ref{tab:nonuniform stratification} for the nonuniform stratification.} FFT ($\widehat{W}$) of  the vertical velocity time series from $t = 34T$ to $t = 40T$ at (a) x = D/4, (b) x = D/2 and (c) x = 3D/4 . In each of the plots, the three curves correspond to $z = H/4, H/2$ and $3H/4$. The solid and dashed vertical lines denote the frequency bands that are used to filter the wave field at $\Omega = \omega_0$ and $\Omega = 2\omega_0$, respectively.} \label{nu_fft}
\end{figure}

Figure \ref{nu_fft} shows the steady state Fourier spectrum of the vertical velocity time series from different spatial locations in the numerical simulation of the resonant case 5 in the nonuniform stratification. The occurrence of the superharmonic frequency $2\omega_0$ is evident at all spatial locations, with an increase in its strength with $x$. Unlike uniform stratification cases, there are no noticeable peaks at frequencies away from $\omega_0$ and $2\omega_0$, probably owing to the relatively smaller forcing amplitudes we have used for the nonuniform stratification. The vertical lines in figure \ref{nu_fft} indicate the frequency bands used for filtering the wave fields at the primary frequency $\omega_0$ and the superharmonic frequency $2\omega_0$.

\begin{figure}
\centering
\includegraphics[width=1\textwidth]{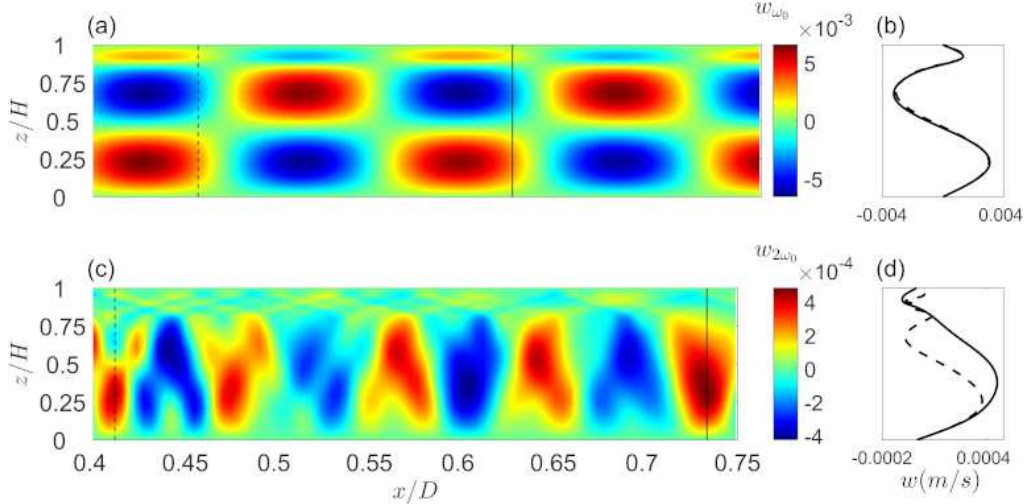}
\caption{{\it Filtered vertical velocity fields from the numerical simulation of case-5 in table \ref{tab:nonuniform stratification} for the nonuniform stratification.} (a) Vertical velocity field $w_{\omega_0}$ at $t = 34T$, after filtering the raw time series during $t = 34T-40T$ at $\Omega = \omega_0$. The bandwidth of frequency used for the filtering is indicated by the vertical solid lines in figure \ref{un_fft}. (b) Vertical profiles of $w_{\omega_0}$ at the two spatial locations corresponding to the vertical dashed and solid lines in (a).  (c) Vertical velocity field $w_{2\omega_0}$ at $t = 34T$, after filtering the raw time series during $t = 34T-40T$ at $\Omega = 2\omega_0$. The bandwidth of frequency used for the filtering is indicated by the vertical dashed lines in figure \ref{nu_fft}. (d) Vertical profiles of $w_{2\omega_0}$ at the two spatial locations corresponding to the vertical dashed and solid lines in (c).}\label{nu_filt}
\end{figure}

Figure \ref{nu_filt} shows the spatial structure of the vertical velocity wave field filtered at $\omega_0$ (top row) and $2\omega_0$ (bottom row). Figure \ref{nu_filt}(a) shows a mode-3 vertical structure with a horizontal wavelength of $\approx7$km, consistent with the forcing provided at $x = 0$. Vertical profiles from two different $x$ locations confirm the mode-3 structure, with no noticeable difference in the amplitude between the two locations (figure \ref{nu_filt}b). The wave field at $2\omega_0$ (figure \ref{nu_filt}c) shows a predominantly mode-2 spatial structure, with the contamination by other modes becoming smaller as we go further away from the forcing location. The vertical profiles shown in 
figure \ref{nu_filt}(d) confirm the predominant mode-2 structure, with a clear increase in the amplitude with $x$. In summary, figure 15 presents a clear evidence from numerical simulations that a superharmonic mode-2 is spontaneously generated as a result of self-interaction of mode-3 at $\omega/N_0 = 0.4466$. We proceed to quantify the modal amplitudes in the wave fields shown in figure \ref{nu_filt}, and quantitatively compare the growth of the superharmonic wave with the prediction from the amplitude evolution equations derived in section \ref{sec:AEE}.

\begin{figure}
\centering
\includegraphics[width=1\textwidth]{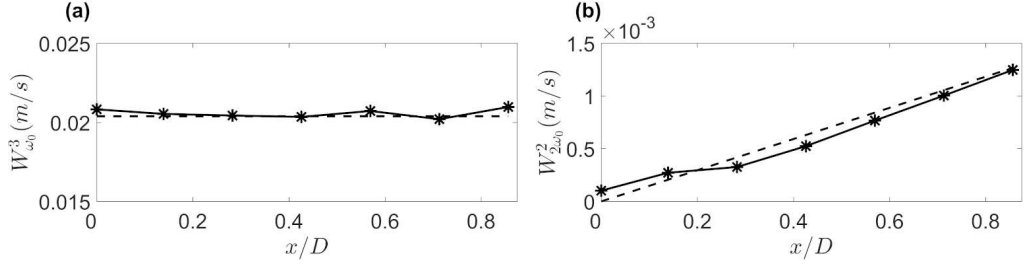}
\caption{Variation with $x$ of (a) mode-3 amplitude at $\Omega = \omega_0$ based on the filtered vertical velocity field shown in figure \ref{nu_filt}(a) from the numerical simulation of case-5 in table \ref{tab:nonuniform stratification} for the nonuniform stratification. The dashed line in (a) indicates the corresponding forcing amplitude at $x=0$. (b) Variation with $x$ of mode-2 amplitude at $\Omega = 2\omega_0$ based on the filtered vertical velocity field shown in figure \ref{nu_filt}(c). The dashed line in (b) represents the linear growth of $W^{2}_{2\omega_0}$ as predicted by the amplitude evolution equations assuming the primary wave amplitude $W^{3}_{\omega_0}$ to be constant at the value shown in (a).}\label{nus_2w}
\end{figure}

Figure \ref{nus_2w}(a) shows the amplitude of mode-3 at $\omega_0$, estimated from the wave field in figure \ref{nu_filt}(a) using the methods described in section \ref{sec:numerical_analysis}. The vertical velocity modal amplitude $W^{3}_{\omega_0}$ is reasonably constant throughout the computational domain, consistent with the forcing amplitude indicated by the dashed line in figure \ref{nus_2w}(a). The spatial growth of the superharmonic mode-2  amplitude ($W^{2}_{2\omega_0}$) is captured accurately by the linear growth predicted by the amplitude evolution equations assuming $W^3_{\omega_0}$ to be constant at the forcing value (figure \ref{nus_2w}b). Thus, figure \ref{nus_2w} provides quantitative validation of the early spatial growth of the superharmonic wave predicted by the amplitude evolution equations derived in section \ref{sec:AEE}. Finally, to verify that the superharmonic wave growth in the numerical simulations is independent of the forcing amplitude or the computational domain length, we present results from two other runs at the resonant frequency in figure \ref{nu_amp} (appendix A). 

\begin{figure}
\centering
\includegraphics[width=1\textwidth]{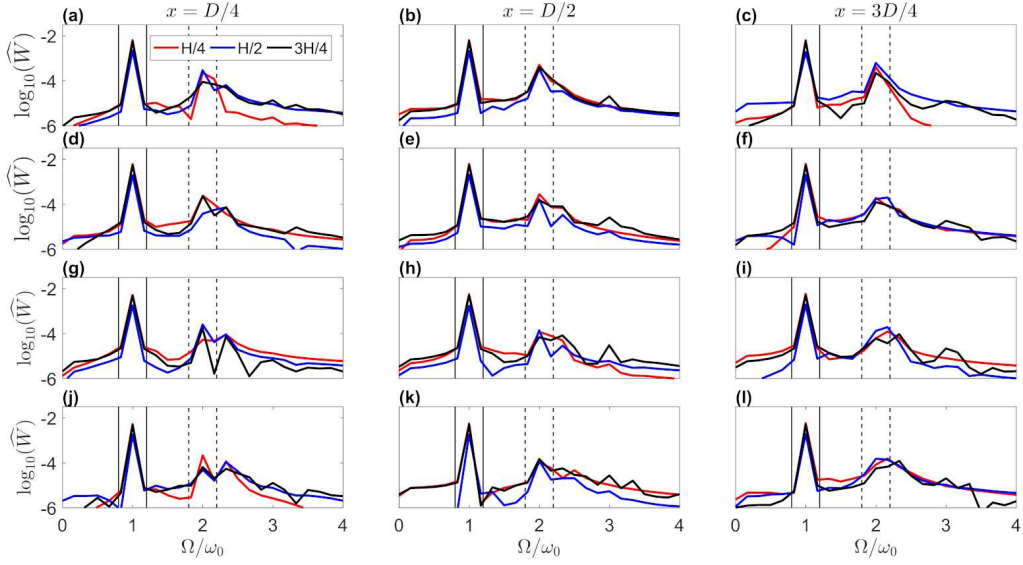}
\caption{{\it Fourier spectra of the vertical velocity at specific spatial locations from the numerical simulations of the off-resonance cases: case-4 (top row),  case-3 (second row), case-2 (thrid row) and case-1 (bottom row) in table \ref{tab:nonuniform stratification} for the nonuniform stratification}. FFT($\widehat{W}$) of the vertical velocity time series from t = 34T to t = 40T at x = D/4 (left column), x = D/2 (middle column) and x = 3D/4 (right column). In each of the plots, the three curves correspond to z = H/4,H/2 and 3H/4. The solid and dashed vertical lines denote the frequency bands that are used to filter the wave field at  $\Omega = \omega_0$ and $\Omega = 2\omega_0$, respectively.}\label{nu_offres}
\end{figure}

\begin{figure}
\centering
\includegraphics[width=1\textwidth]{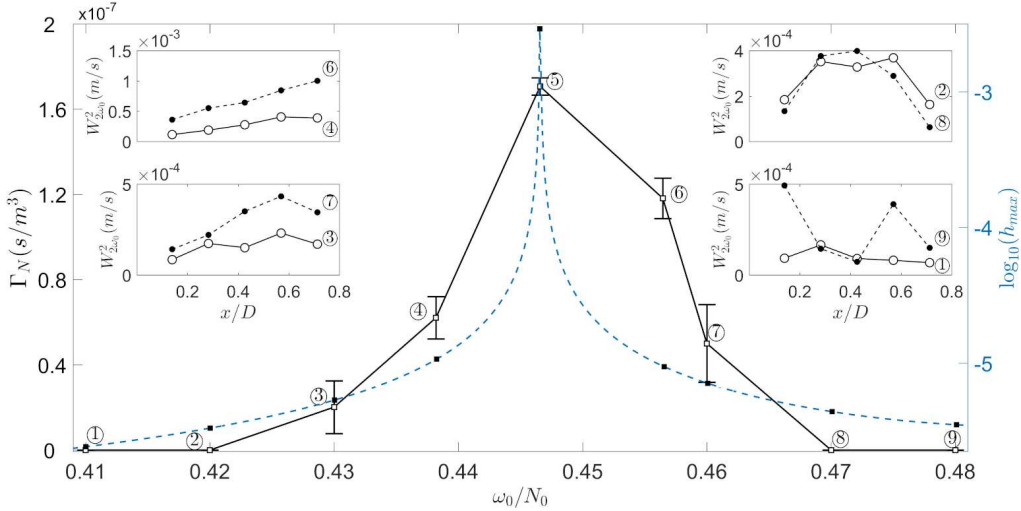}
\caption{The normalized initial growth rate $\Gamma_N = d|\Psi_3|/dX/(|\Psi_1(X=0)||\Psi_2(X=0)|)$ of mode-2 at $\Omega = 2\omega_0$, estimated from the variation of $W_{2\omega_0}^2$ with $x$, plotted as a function of $\omega_0/N_0$ based on numerical simulations of all the cases in table \ref{tab:nonuniform stratification} for the nonuniform stratification. The encircled numbers next to every data point represent the corresponding case numbers. The insets show the $W_{2\omega_0}^2$ vs. $x/D$ plots for cases 1-9, except the resonant case-5 whose variation is shown in figure 9. The blue dashed curve (with the black solid markers) shows the variation of the amplitude of the superharmonic mode-1 ($log_{10}(h_{max})$, plotted on the right axis) based on the steady-state theory presented in section \S~\ref{offres}.}\label{nu_grand}
\end{figure}

While the amplitude evolution equations derived in section \ref{sec:AEE} are valid only at the resonant frequency, it is worthwhile investigating the generation of superharmonic waves at off-resonant frequencies as well. Towards this objective, we ran eight more numerical simulations, as described by cases 1-4 and 6-9 listed in table \ref{tab:nonuniform stratification}. Figure \ref{nu_offres} shows the steady state Fourier spectra of the vertical velocity time series at different spatial locations as we move away from the resonant frequency, i.e. cases 4 to 1 from the top to bottom rows. The occurrence of a spatially growing superharmonic wave is evident for case-4 (top row of figure \ref{nu_offres}), which is closest to the resonant case-5. At other off-resonant frequencies, relatively weaker strengths are observed at $2\omega_0$, while newer frequencies higher than $2\omega_0$ seem to emerge. We quantify the spatial growth of the superharmonic mode-2 at various off-resonant primary wave forcing frequencies in figure \ref{nu_grand}, whereas a detailed investigation of the generation of even higher frequencies ($>2\omega_0$) is left out of the current study as it is outside the scope of the amplitude evolution equations derived in section \ref{sec:AEE}.

As mentioned in \S~\ref{sec:numerical_analysis}, the flow fields filtered at 2$\omega_0$ from the off-resonant cases were subject to modal decomposition too, though the superharmonic waves are not necessarily internal waves. Figure \ref{nu_grand} shows the variation of the normalized initial growth rate $\Gamma_N$ ($= \Gamma/(|\Psi_1(X=0)||\Psi_2(X=0)|$) of the supherharmonic mode-2 amplitude with $\omega_0/N_0$. As expected, the value of $\Gamma_N$ is $\alpha_3$ at the resonant case-5, and decreases towards zero as we go far from the resonant frequency. Interestingly, there is noticeable generation of the superharmonic mode-2 even in cases 4 and 6, which are close to but not at resonance. Finally, the width of the peak in $\Gamma_N$ vs. $\omega_0$ seems wider than the peak in the variation of the amplitude from steady-state theory, suggesting that the superharmonic wave generation due to triadic resonance is not restricted to the resonant frequency alone.   

\section{Discussion and Conclusions}
\label{sec:disc}

In this paper, we have derived the amplitude evolution equations that govern energy transfer between three internal wave modes that form a resonant triad in a finite-depth fluid with an arbitrary stratification profile and background rotation. Allowing slow spatial evolution of the three modal amplitudes, and invoking the triadic resonance condition in a nonuniform stratification \citep{JFM2017}, the amplitude evolution equations essentially ensure that resonant divergence of the weakly nonlinear solution does not occur even when finite energy is present in all the waves that form the resonant triad. Our theoretical framework could potentially model how the energy in a linearly forced internal wave field such as the internal tides or near-inertial internal waves could get redistributed to other frequencies and wavenumbers as the waves propagate away from their generation location. It is noteworthy that the amplitude evolution equations derived in this paper are valid for any internal wave resonant triad containing modes, though the specific cases studied in \S~\ref{subsec:uniform} and \S~\ref{subsec:nonuniform} focused on superharmonic wave generation due to modal interactions. The primary wave energy decay due to superharmonic wave generation could potentially be significant since an appreciable amount of energy is initially injected into two of the waves that form the superharmonic resonant triad.   

In section \S~\ref{subsec:uniform}, we considered the scenario of modes 1 \& 2 at $\omega/N_0 = 0.3953$ in triadic resonance with mode-1 at the superharmonic frequency $2\omega$ in a uniform stratification. The numerical solutions of the amplitude evolution equations for this resonant triad showed that a significant fraction of the primary wave energy would be transferred to the superharmonic wave for realistic ocean-like parameters. Direct numerical simulations (DNS) then provided evidence for the spontaneous generation of the superharmonic mode-1 as a result of resonant interaction between modes 1 \& 2 at $\omega_0/N_0 = 0.3953$. The initial spatial growth rate predicted by the amplitude evolution equations were quantitatively validated by the DNS. Furthermore, DNS were run at off-resonant primary wave frequencies to show that superharmonic wave generation occurs away from the triadic resonance as well. In \S~\ref{nu_results}, we studied the case of a self-interacting mode-3 at $\omega_0/N_0 = 0.4466$ in triadic resonance with the superharmonic mode-2 in a nonuniform stratification with a pycnocline. The spontaneous generation of the superharmonic mode-2 even when it has no initial energy was demonstrated in DNS, which was also used to quantitatively validate the initial spatial growth predicted by the amplitude evolution equations. Finally, like for the uniform stratification, superharmonic wave generation is observed even at off-resonant frequencies in the nonuniform stratification. Finally, the specific triadic resonances considered in section \ref{sec:results} are only representative, and similar studies involving superharmonic wave generation due to higher mode interactions could be useful. 

\begin{figure}
\centering
\includegraphics[width=1\textwidth]{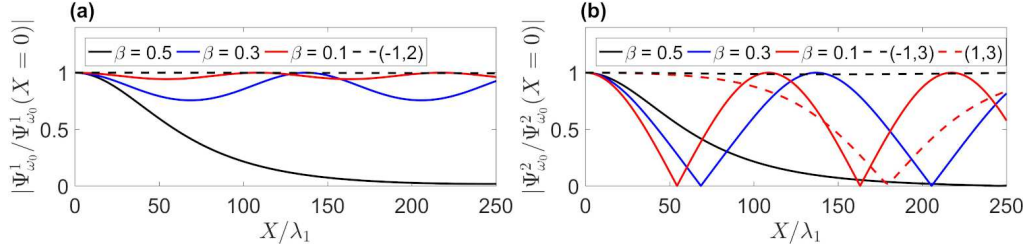}
\caption{Spatial evolution of the primary wave amplitudes: (a) mode-1 at frequency $\omega_0/N_0 = 0.3953$, (b) mode-2 at frequency $\omega_0/N_0 = 0.3953$, as obtained from the numerical solutions of the amplitude evolution equations corresponding to specific superharmonic and subharmonic resonant triads comprising the primary waves in a uniform stratification. Specifically, the solid curves in both (a) and (b) correspond to the superharmonic resonant triad discussed in section \ref{ut_results}, with the value of $\beta$ indicated in the legend. The dashed curves correspond to subharmonic resonant triads, with the secondary waves indicated by the mode numbers in the legend, and the primary wave being (a) mode-1 at $\omega_0/N_0 = 0.3953$ and (b) mode-2 at $\omega_0/N_0 = 0.3953$. A negative mode number indicates propagation in a direction opposite to the primary waves. For the subharmonic resonant triads, the daughter waves were initiated with an amplitude that corresponds to 1\% of the primary wave energy. For the superharmonic resonant triads, $E_T = 2000$W/m has been assumed. }\label{sub_sup}
\end{figure}

Apart from the superharmonic resonant triad, each of the primary waves is also a part of infinitely many subharmonic resonant triads. For example, the left-to-right propagating mode-1 at $\omega_0/N_0 = 0.3953$ in the uniform stratification is also in triadic resonance with the right-to-left propagating mode-1 at $\omega_1/N_0 = 0.1266$ and the left-to-right propagating mode-2 at $\omega_2/N_0 = 0.2687$. As shown in figure \ref{sub_sup}(a), this subharmonic resonant triad is very inefficient in transferring energy out of the primary wave to the subharmonic secondary waves, which typically have a very small amount of initial energy that is available from the background noise. In contrast, the superharmonic resonant triad can transfer a significant fraction of the primary mode-1 energy to the superharmonic wave, with the fraction being larger for larger $\beta$ and independent of $E_T$. Interestingly, there exists no subharmonic resonant triad for the mode-1 primary wave at $\omega_0/N_0 = 0.3953$ such that both the daughter waves propagate in the same direction as the primary wave. In figure \ref{sub_sup}(b), we provide a comparison between the superharmonic resonant triad and two representative subharmonic resonant triads that comprise mode-2 at $\omega_0/N_0 = 0.3953$ as the primary wave. The subharmonic resonant triad with mode-1 at $\omega_1/N_0 = 0.1725$ and mode-3 at $\omega_2/N_0 = 0.2228$ as daughter waves is observed to be capable of extracting all the energy out of the primary mode-2, but at a rate that is significantly slower than the superharmonic resonant triad for all three values of $\beta$ considered. For the primary wave decay rate in the superharmonic resonant triad with $\beta = 0.1$ to become comparable to what is observed for the (1,3) subharmonic resonant triad in figure \ref{sub_sup}(b), $E_T$ has to be reduced by a factor of around 9 (recall from section \ref{ut_results} that the normalized decay rate scales as $1/\sqrt{E_T}$). In other words, even for $E_T$ as small as around 200W/m, the primary wave decay rate due to the superharmonic resonant triad is comparable to what occurs due to the (1,3) subharmonic resonant triad. In summary, figure \ref{sub_sup} suggests that the superharmonic wave generation is likely to significantly modify any energy transfer estimates due to triadic resonances, at least in and around the resonant primary wave frequencies. Future studies that account for all possible subharmonic resonant triads, and their simultaneous evolution with the superharmonic resonant triad would be insightful.        

The real ocean scenario typically has several modes excited simultaneously at the forcing frequency. With respect to internal tides, it is well recognized that a barotropic forcing on the ocean floor topography excites several modes at the tidal frequency, with the lower modes carrying relatively larger amount of energy. Similarly, wind forcing on the ocean surface is known to input energy into various modes at the near-inertial frequency. Furthermore, \cite{JFM2017} have shown that the occurrence of a resonant triad containing two modes at the forcing frequency, and a third mode at the superharmonic frequency is highly likely when more than a few modes are simultaneously generated. It therefore seems important that superharmonic wave generation as a result of modal interaction is considered alongside other triadic resonances like PSI when a finite amount of energy is put into several modes at a given forcing frequency. Future theoretical studies could incorporate the simultaneous presence of several modes at the forcing frequency, and all the resonant triads that they are a part of. It may also be worthwhile to investigate the role of background noise that may inject small but finite amounts of energy in the resonant triad components that are not at the forcing frequency. Finally, incorporating viscous dissipation in the amplitude evolution equations would make them more relevant to describe observations from laboratory/numerical/ field experiments.  

Direct numerical simulations and/or laboratory experiments could be used to further validate the amplitude evolution equations over a wider range of scenarios than what is considered in this paper. Amplitude evolution over spatially large extents where the growth of the superharmonic wave would not be linear would be particularly useful to investigate. Further numerical or experimental studies at off-resonant frequencies should shed light on the role of superharmonic wave generation in the ensuing instabilities, the transition to turbulence and subsequent dissipation of internal waves. Finally, our study shows that superharmonic wave generation in internal tides or near-inertial internal waves may not be uncommon, and our theoretical framework could provide an understanding of results from DNS or laboratory experiments or field studies. For example, the superharmonic wave generation upon wave beam reflection at a pycnocline \citep{diamessis_etal,mercier_etal} could potentially be described using the ideas developed in this paper.  

\begin{acknowledgments}
The authors thank the Ministry of Earth Sciences, Government of India and the Department of Science \& Technology, Government of India for financial support under the Monsoon Mission Grant MM/2014/IND-002 and the FIST grant SR/FST/ET-II/2017/109, respectively. Fruitful discussions with Eric d'Asaro are acknowledged. The manuscript was partly written during the stay of M.M. at Laboratoire de Physique, ENS de Lyon, and their hospitality is gratefully acknowledged.
\end{acknowledgments}

\appendix

\section{Governing equations at $\mathcal{O} (\epsilon^0)$, $\mathcal{O} (\epsilon^1)$ and $\mathcal{O} (\epsilon^2)$}
\label{eps_eqn}

The governing equations (\ref{psi_eqn}) - (\ref{v_eqn}) at $\mathcal{O} (\epsilon^0)$ are
\begin{gather}
\frac{\partial^2}{\partial t^2}(\nabla^{2}\psi_{0}) + f^{2}\frac{\partial^{2}\psi_{0}}{\partial z^{2}} = \frac{g}{\bar{\rho}}\frac{\partial}{\partial x}[J(\psi_0,\rho_0)] - \frac{\partial}{\partial t}[J(\psi_0,\nabla^{2}\psi_0)] + f \frac{\partial}{\partial z}[J(\psi_0,v_0)],\nonumber \\
\frac{\partial\rho_0}{\partial t} = -J(\psi_0,\rho_0),\nonumber\\
\frac{\partial v_0}{\partial t} - f \frac{\partial \psi_0}{\partial z} = - J(\psi_0,v_0).\label{eps0} 
\end{gather}

The governing equations (\ref{psi_eqn}) - (\ref{v_eqn}) at $\mathcal{O} (\epsilon^1)$ are 
\begin{align}
\frac{\partial^2}{\partial t^2}&(\nabla^{2}\psi_{1}) + 2\frac{\partial^{2}}{\partial t^2 }\bigg(\frac{\partial^{2}\psi_{0}}{\partial x \partial X}\bigg) + f^{2}\frac{\partial^{2}\psi_{1}}{\partial z^{2}}=\nonumber \\
& \frac{g}{\bar{\rho}}\frac{\partial}{\partial x}[J(\psi_0,\rho_1) + J(\psi_1,\rho_0)]+ \frac{g}{\bar{\rho}}\bigg[\frac{\partial}{\partial x}[J_{X}(\psi_0,\rho_0)] + \frac{\partial}{\partial X}[J(\psi_0,\rho_0)]\bigg]\nonumber \\
- &\frac{\partial}{\partial t}[J(\psi_0,\nabla^{2}\psi_1) + J(\psi_1,\nabla^{2}\psi_0)]  - \frac{\partial}{\partial t}[J_{X}(\psi_0,\nabla^{2}\psi_0) + 2J(\psi_0,\frac{\partial^{2}\psi_0}{\partial x \partial X})] \nonumber \\
+& f \frac{\partial}{\partial z}[J(\psi_0,v_1) + J(\psi_1,v_0) + J_{X}(\psi_0,v_0)], \nonumber
\end{align}
\begin{align} 
&\frac{\partial \rho_{1}}{\partial t}  =  -[J(\psi_0,\rho_1) + J(\psi_1,\rho_0) + J_{X}(\psi_0,\rho_0)],\nonumber \\  
&\frac{\partial v_1}{\partial t}  -  f \frac{\partial \psi_1}{\partial z} = -[ J(\psi_0,v_1) + J(\psi_1,v_0) + J_{X}(\psi_0,v_0)] .\label{eps1}
\end{align}

The governing equations (\ref{psi_eqn}) - (\ref{v_eqn}) at $\mathcal{O} (\epsilon^2)$ are 
\begin{align*}
\frac{\partial^2}{\partial t^2}&(\nabla^{2}\psi_{2}) + 2\frac{\partial^{2}}{\partial t^2 }\bigg(\frac{\partial^{2}\psi_{1}}{\partial x \partial X}\bigg) + \frac{\partial^{2}}{\partial t^2}\bigg(\frac{\partial^{2}\psi_0}{\partial X^{2}}\bigg) + f^{2} \frac{\partial^{2}\psi_{2}}{\partial z^{2}}=\nonumber \\
&\frac{g}{\bar{\rho}}\frac{\partial}{\partial x}[J(\psi_0,\rho_2) + J(\psi_1,\rho_1) + J(\psi_2,\rho_0)]\nonumber \\
+& \frac{g}{\bar{\rho}}\bigg[\frac{\partial}{\partial x}[J_{X}(\psi_0,\rho_1) + J_{X}(\psi_1,\rho_0)] + \frac{\partial}{\partial X}[J(\psi_0,\rho_1) + J(\psi_1,\rho_0)]\bigg] \nonumber \\
+& \frac{g}{\bar{\rho}}\frac{\partial}{\partial X}[J_{X}(\psi_0,\rho_0)] - \frac{\partial}{\partial t}[J(\psi_0,\nabla^{2}\psi_2) + J(\psi_1,\nabla^{2}\psi_1) + J(\psi_2,\nabla^{2}\psi_0)] \nonumber \\
-&\frac{\partial}{\partial t}[J_{X}(\psi_0,\nabla^{2}\psi_1) + J_{X}(\psi_1,\nabla^{2}\psi_0) + 2J(\psi_0,\frac{\partial^{2}\psi_1}{\partial x \partial X}) + 2J(\psi_1,\frac{\partial^{2}\psi_0}{\partial x \partial X})]\nonumber\\ 
-&\frac{\partial}{\partial t}[J(\psi_0,\frac{\partial^{2}\psi_0}{\partial X^2}) + 2J_{X}(\psi_0,\frac{\partial^{2}\psi_0}{\partial x \partial X})]  \nonumber \\
+ &f \frac{\partial}{\partial z}[J(\psi_0,v_2) + J(\psi_1,v_1) + J(\psi_2,v_0)] + f\frac{\partial}{\partial z}[J_{X}(\psi_0,v_1) + J_{X}(\psi_1,v_0)] , \nonumber
\end{align*}
\begin{align}
&\frac{\partial \rho_{2}}{\partial t} 
 = - [(J(\psi_0,\rho_2) + J(\psi_1,\rho_1) + J(\psi_2,\rho_0) + J_{X}(\psi_0,\rho_1) + J_{X}(\psi_1,\rho_0)],\nonumber\\
&\frac{\partial v_2}{\partial t} - f \frac{\partial \psi_2}{\partial z}
 = -[ J(\psi_0,v_2) + J(\psi_1,v_1) + J(\psi_2,v_0) + J_{X}(\psi_0,v_1) + J_{X}(\psi_1,v_0)].\nonumber\\ \label{eps2}
\end{align}
In equations (\ref{eps1}) and (\ref{eps2}), the operator $J_X$ is given by $J_{X}(A,B) = (\partial A/\partial X)(\partial B/\partial z) - (\partial B/\partial X)(\partial A/\partial z)$.

\section{Numerical simulations of resonant cases: Convergence study}
\label{convergence}

\begin{figure}
\centering
\includegraphics[width=1\textwidth]{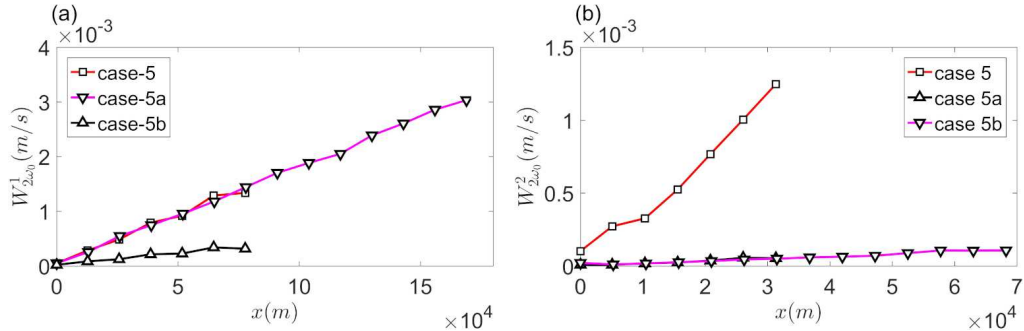}
\caption{Variation with $x$ of the amplitude of the superharmonic secondary wave for (a) cases 5, 5a and 5b in table \ref{tab:Uniform stratification} for the uniform stratification, and (b) cases 5, 5a and 5b in table \ref{tab:nonuniform stratification} for the nonuniform stratification.}\label{nu_amp}
\end{figure}

To establish the numerical convergence of the spatial growth of the superharmonic wave in the DNS of resonant cases in uniform and nonuniform stratifications, we ran the same resonant cases with smaller forcing amplitude and/or longer computational domain. As detailed in table \ref{tab:Uniform stratification} for the uniform stratification, case 5a corresponds to a longer computational domain but at the same forcing amplitudes as case 5. Case 5b corresponds to the same computational domain as for case 5, but with half the forcing amplitudes as that of case 5. Similarly, cases 5, 5a and 5b in figure \ref{nu_amp}(b) are listed in table \ref{tab:nonuniform stratification} for the nonuniform stratification. While case 5a corresponds to one-fifths forcing amplitude as that of case 5, case 5b also has a longer computational domain. The results in figure \ref{nu_amp} suggest that the initial growth rates observed in the DNS are the same for the two computational domain lengths considered, and scale with the forcing amplitudes in a manner consistent with the amplitude evolution equations derived in section \ref{sec:AEE}.  

\bibliographystyle{jfm}
\bibliography{bib_file1}

\begin{thebibliography}{45}
\expandafter\ifx\csname natexlab\endcsname\relax\def\natexlab#1{#1}\fi

\bibitem[Alford(2003{\natexlab{{\em a\/}}})]{alford2003}
{\sc Alford, Matthew~H} 2003{\natexlab{{\em a\/}}} Improved global maps and
  54-year history of wind-work on ocean inertial motions. {\em Geophysical
  Research Letters\/} {\bf 30}~(8).

\bibitem[Alford(2003{\natexlab{{\em b\/}}})]{alford03}
{\sc Alford, Matthew~H} 2003{\natexlab{{\em b\/}}} Redistribution of energy
  available for ocean mixing by long-range propagation of internal waves. {\em
  Nature\/} {\bf 423}~(6936), 159.

\bibitem[Alford {\em et~al.\/}(2016)Alford, MacKinnon, Simmons \&
  Nash]{alford2016}
{\sc Alford, Matthew~H, MacKinnon, Jennifer~A, Simmons, Harper~L \& Nash,
  Jonathan~D} 2016 Near-inertial internal gravity waves in the ocean. {\em
  Annual review of marine science\/} {\bf 8}, 95--123.

\bibitem[Benielli \& Sommeria(1998)]{benielli1998}
{\sc Benielli, Dominique \& Sommeria, Joel} 1998 Excitation and breaking of
  internal gravity waves by parametric instability. {\em Journal of Fluid
  Mechanics\/} {\bf 374}, 117--144.

\bibitem[Bourget {\em et~al.\/}(2013)Bourget, Dauxois, Joubaud \&
  Odier]{Bourget_etal}
{\sc Bourget, B., Dauxois, T., Joubaud, S. \& Odier, P.} 2013 Experimental
  study of parametric subharmonic instability for internal plane waves. {\em J.
  Fluid Mech.\/} {\bf 723}, 1--20.

\bibitem[Bourget {\em et~al.\/}(2014)Bourget, Scolan, Dauxois, Le~Bars, Odier
  \& Joubaud]{bourget2014}
{\sc Bourget, Baptiste, Scolan, H{\'e}l{\`e}ne, Dauxois, Thierry, Le~Bars,
  Michael, Odier, Philippe \& Joubaud, Sylvain} 2014 Finite-size effects in
  parametric subharmonic instability. {\em Journal of Fluid Mechanics\/} {\bf
  759}, 739--750.

\bibitem[Brucker \& Sarkar(2010)]{sponge_ref}
{\sc Brucker, Kyle~A \& Sarkar, Sutanu} 2010 A comparative study of
  self-propelled and towed wakes in a stratified fluid. {\em J. Fluid Mech.\/}
  {\bf 652}, 373--404.

\bibitem[Chalamalla {\em et~al.\/}(2017)Chalamalla, Santilli, Scotti, Jalali \&
  Sarkar]{vamsi2017}
{\sc Chalamalla, Vamsi~K, Santilli, Edward, Scotti, Alberto, Jalali, Masoud \&
  Sarkar, Sutanu} 2017 Somar-les: A framework for multi-scale modeling of
  turbulent stratified oceanic flows. {\em Ocean Modelling\/} {\bf 120},
  101--119.

\bibitem[Colella \& Woodward(1984)]{Colella1984}
{\sc Colella, Phillip \& Woodward, Paul~R} 1984 The piecewise parabolic method
  (ppm) for gas-dynamical simulations. {\em Journal of computational physics\/}
  {\bf 54}~(1), 174--201.

\bibitem[Dauxois {\em et~al.\/}(2018)Dauxois, Joubaud, Odier \&
  Venaille]{dauxois}
{\sc Dauxois, Thierry, Joubaud, Sylvain, Odier, Philippe \& Venaille, Antoine}
  2018 Instabilities of internal gravity wave beams. {\em Annual Review of
  Fluid Mechanics\/} {\bf 50}, 131--156.

\bibitem[Diamessis {\em et~al.\/}(2014)Diamessis, Wunsch, Delwiche \&
  Richter]{diamessis_etal}
{\sc Diamessis, P.J., Wunsch, S., Delwiche, I. \& Richter, M.P.} 2014 Nonlinear
  generation of harmonics through the interaction of an internal wave beam with
  a model oceanic pycnocline. {\em Dynam. Atmos. Oceans.\/} {\bf 66}, 110--137.

\bibitem[Egbert \& Ray(2000)]{egbertray}
{\sc Egbert, GD \& Ray, RD} 2000 Significant dissipation of tidal energy in the
  deep ocean inferred from satellite altimeter data. {\em Nature\/} {\bf
  405}~(6788), 775.

\bibitem[Garrett(2003)]{garrett2003}
{\sc Garrett, Chris} 2003 Internal tides and ocean mixing. {\em Science\/} {\bf
  301}~(5641), 1858--1859.

\bibitem[Garrett \& Kunze(2007)]{GarrettKunze07}
{\sc Garrett, C. \& Kunze, E.} 2007 Internal tide generation in the deep ocean.
  {\em Annu. Rev. Fluid Mech.\/} {\bf 39}, 57--87.

\bibitem[Garrett \& Munk(1979)]{garrett_munk}
{\sc Garrett, Christopher \& Munk, Walter} 1979 Internal waves in the ocean.
  {\em Annual review of fluid mechanics\/} {\bf 11}~(1), 339--369.

\bibitem[Gayen \& Sarkar(2013)]{gayen_sarkar}
{\sc Gayen, B. \& Sarkar, S.} 2013 Degradation of an internal wave beam by
  parametric subharmonic instability in an upper ocean pycnocline. {\em J.
  Geophys. Res.\/} {\bf 118}, 4689--4698.

\bibitem[Ghaemsaidi \& Mathur(2019)]{ghaemsaidi2019}
{\sc Ghaemsaidi, Sasan~John \& Mathur, Manikandan} 2019 Three-dimensional
  small-scale instabilities of plane internal gravity waves. {\em Journal of
  Fluid Mechanics\/} {\bf 863}, 702--729.

\bibitem[Gill(1984)]{gill}
{\sc Gill, A.E.} 1984 On the behavior of internal waves in the wakes of storms.
  {\em J. Phys. Oceanogr.\/} {\bf 14}~(7), 1129--1151.

\bibitem[Joubaud {\em et~al.\/}(2012)Joubaud, Munroe, Odier \&
  Dauxois]{Joubaud_etal}
{\sc Joubaud, S., Munroe, J., Odier, P. \& Dauxois, T.} 2012 Experimental
  parametric subharmonic instability in stratified fluids. {\em Phys. Fluids\/}
  {\bf 24}, 041703.

\bibitem[Karimi \& Akylas(2014)]{karimi_akylas}
{\sc Karimi, H.H. \& Akylas, T.R.} 2014 Parametric subharmonic instability of
  internal waves: locally confined beams versus monochromatic wavetrains. {\em
  J. Fluid Mech.\/} {\bf 757}, 381--402.

\bibitem[Karimi \& Akylas(2017)]{karimi2017}
{\sc Karimi, Hussain~H \& Akylas, TR} 2017 Near-inertial parametric subharmonic
  instability of internal wave beams. {\em Physical Review Fluids\/} {\bf
  2}~(7), 074801.

\bibitem[Klostermeyer(1991)]{klostermeyer1991}
{\sc Klostermeyer, J} 1991 Two-and three-dimensional parametric instabilities
  in finite-amplitude internal gravity waves. {\em Geophysical \& Astrophysical
  Fluid Dynamics\/} {\bf 61}~(1-4), 1--25.

\bibitem[Koudella \& Staquet(2006)]{Koudella_Staquet}
{\sc Koudella, C.~R. \& Staquet, C.} 2006 Instability mechanisms of a
  two-dimensional progressive internal gravity wave. {\em J. Fluid Mech.\/}
  {\bf 548}, 165--196.

\bibitem[LeBlond \& Mysak(1981)]{leblond}
{\sc LeBlond, P.H. \& Mysak, L.A.} 1981 {\em Waves in the Ocean\/}. Elsevier.

\bibitem[Van~der Lee \& Umlauf(2011)]{van2011}
{\sc Van~der Lee, EM \& Umlauf, L} 2011 Internal wave mixing in the baltic sea:
  Near-inertial waves in the absence of tides. {\em Journal of Geophysical
  Research: Oceans\/} {\bf 116}~(C10).

\bibitem[Liang {\em et~al.\/}(2017)Liang, Couston, Guo \& Alam]{liang2017}
{\sc Liang, Yong, Couston, Louis-Alexandre, Guo, Qiuchen \& Alam,
  Mohammad-Reza} 2017 Dominant resonance in parametric subharmonic instability
  of internal waves. {\em arXiv preprint arXiv:1709.06250\/} .

\bibitem[MacKinnon \& Winters(2005)]{mackinnon_winters}
{\sc MacKinnon, J.~A. \& Winters, K.~B.} 2005 Subtropical catastrophe:
  Significant loss of low-mode tidal energy at 28.9. {\em Geophys. Res.
  Lett.\/} {\bf 32}~(15).

\bibitem[Martin {\em et~al.\/}(1972)Martin, Simmons \& Wunsch]{martin}
{\sc Martin, S., Simmons, W. \& Wunsch, C.} 1972 The excitation of resonant
  triads by single internal waves. {\em J. Fluid Mech.\/} {\bf 53}~(01),
  17--44.

\bibitem[McEwan(1971)]{mcewan}
{\sc McEwan, A.D.} 1971 Degeneration of resonantly-excited standing internal
  gravity waves. {\em J. Fluid Mech.\/} {\bf 50}~(03), 431--448.

\bibitem[McEwan {\em et~al.\/}(1972)McEwan, Mander \& Smith]{mcewan1972}
{\sc McEwan, AD, Mander, DW \& Smith, RK} 1972 Forced resonant second-order
  interaction between damped internal waves. {\em Journal of Fluid Mechanics\/}
  {\bf 55}~(4), 589--608.

\bibitem[Mercier {\em et~al.\/}(2012)Mercier, Mathur, Gostiaux, Gerkema,
  Magalhaes, Da~Dilva \& Dauxois]{mercier_etal}
{\sc Mercier, M., Mathur, M., Gostiaux, L., Gerkema, T., Magalhaes, J.M.,
  Da~Dilva, J.C.B. \& Dauxois, T.} 2012 Soliton generation by internal tidal
  beams impinging on a pycnocline: laboratory experiments. {\em J. Fluid
  Mech.\/} {\bf 704}, 37--60.

\bibitem[Munk \& Wunsch(1998)]{MunkWunsch98}
{\sc Munk, W. \& Wunsch, C.} 1998 Abyssal recipes ii: energetics of tidal and
  wind mixing. {\em Deep-Sea Res.\/} {\bf 45}, 1977--2010.

\bibitem[Nayfeh(2008)]{nayfeh_book}
{\sc Nayfeh, Ali~H} 2008 {\em Perturbation methods\/}. John Wiley \& Sons.

\bibitem[Pollard(1970)]{pollard1970}
{\sc Pollard, Raymond~T} 1970 On the generation by winds of inertial waves in
  the ocean {\bf 17}~(4), 795--812.

\bibitem[Rainville \& Pinkel(2006)]{rainville2006}
{\sc Rainville, Luc \& Pinkel, Robert} 2006 Propagation of low-mode internal
  waves through the ocean. {\em Journal of Physical Oceanography\/} {\bf
  36}~(6), 1220--1236.

\bibitem[Richet {\em et~al.\/}(2018)Richet, Chomaz \& Muller]{richet2018}
{\sc Richet, O, Chomaz, J-M \& Muller, C} 2018 Internal tide dissipation at
  topography: triadic resonant instability equatorward and evanescent waves
  poleward of the critical latitude. {\em Journal of Geophysical Research:
  Oceans\/} {\bf 123}~(9), 6136--6155.

\bibitem[Santilli \& Scotti(2015)]{SOMAR}
{\sc Santilli, Edward \& Scotti, Alberto} 2015 The stratified ocean model with
  adaptive refinement (somar). {\em Journal of Computational Physics\/} {\bf
  291}, 60--81.

\bibitem[Saranraj \& Guha(2019)]{saranraj}
{\sc Saranraj, G \& Guha, Anirban} 2019 Energy transfer in resonant and
  near-resonant internal wave triads for weakly non-uniform stratifications .

\bibitem[Sonmor \& Klaassen(1997)]{sonmor1997}
{\sc Sonmor, LJ \& Klaassen, GP} 1997 Toward a unified theory of gravity wave
  stability. {\em Journal of the atmospheric sciences\/} {\bf 54}~(22),
  2655--2680.

\bibitem[Staquet \& Sommeria(2002)]{staquet}
{\sc Staquet, C. \& Sommeria, J.} 2002 Internal gravity waves: from
  instabilities to turbulence. {\em Annu. Rev. Fluid Mech.\/} {\bf 34}~(1),
  559--593.

\bibitem[Sutherland(2016)]{sutherland}
{\sc Sutherland, B.R.} 2016 Excitation of superharmonics by internal modes in
  non-uniformly stratified fluid. {\em J. Fluid Mech.\/} {\bf 793}, 335--352.

\bibitem[Thorpe(1966)]{thorpe}
{\sc Thorpe, S.A.} 1966 On wave interactions in a stratified fluid. {\em J.
  Fluid Mech.\/} {\bf 24}~(04), 737--751.

\bibitem[Varma \& Mathur(2017)]{JFM2017}
{\sc Varma, D. \& Mathur, M.} 2017 Internal wave resonant triads in
  finite-depth non-uniform stratifications. {\em J. Fluid Mech.\/} {\bf 824},
  286--311.

\bibitem[Wunsch(2017)]{wunsch2017}
{\sc Wunsch, Scott} 2017 Harmonic generation by nonlinear self-interaction of a
  single internal wave mode. {\em Journal of Fluid Mechanics\/} {\bf 828},
  630--647.

\bibitem[Young {\em et~al.\/}(2008)Young, Tsang \& Balmforth]{young_etal}
{\sc Young, W.R., Tsang, Y.-K. \& Balmforth, N.J.} 2008 Near-inertial
  parametric subharmonic instability. {\em J. Fluid Mech.\/} {\bf 607}, 25--49.

\end{thebibliography}
\end{document}